\documentclass[transmag]{IEEEtran}

\usepackage{latexsym}
\usepackage{amsmath,amssymb,amsfonts}
\usepackage{graphicx}
\usepackage[nolist]{acronym}
\usepackage{cite}
\usepackage{hyperref}

\begin{document}

\title{Feasibility of Modeling Orthogonal Frequency-Division Multiplexing Communication Signals with Unsupervised Generative \\Adversarial Networks}

\author{Jack Sklar and Adam Wunderlich
\thanks{J. Sklar and A. Wunderlich are with the Communications Technology Laboratory, National Institute of Standards and Technology, Boulder, CO, 80305 USA. e-mail: jack.sklar@nist.gov, adam.wunderlich@nist.gov.} %
\thanks{U.S. government work not protected by U.S. copyright.}}

\IEEEtitleabstractindextext{\begin{abstract}
High-quality recordings of radio frequency (RF) emissions from commercial communication hardware in realistic environments are often needed to develop and assess spectrum-sharing technologies and practices, \emph{e.g.}, for training and testing spectrum sensing algorithms and for interference testing.  Unfortunately, the time-consuming, expensive nature of such data collections together with data-sharing restrictions pose significant challenges that limit data set availability.  Furthermore, developing accurate models of real-world RF emissions from first principles is often very difficult because system parameters and implementation details are at best only partially known, and complex system dynamics are difficult to characterize.  Hence, there is a need for flexible, data-driven methods that can leverage existing data sets to synthesize additional similar waveforms.  One promising machine-learning approach is unsupervised deep generative modeling with generative adversarial networks (GANs).  To date, GANs for RF communication signals have not been studied thoroughly.  In this paper, we present the first in-depth investigation of generated signal fidelity for GANs trained with baseband orthogonal frequency-division multiplexing (OFDM) signals, where each subcarrier is digitally modulated with quadrature amplitude modulation (QAM).  Building on prior GAN methods, we developed two novel GAN models and evaluated their performance using simulated data sets with known ground truth.  Specifically, we investigated model performance with respect to increasing data set complexity over a range of OFDM parameters and conditions, including fading channels.  The findings presented here inform the feasibility of use cases and provide a foundation for further investigations into deep generative models for RF communication signals. 
\end{abstract}

}

\maketitle

\section{Introduction}
\label{sec:introduction}

\IEEEPARstart{T}{o} aid the development of spectrum-sharing technologies, practices, and policies \cite{Papadias2020,Voicu2019} for \ac{RF} communication systems, it is necessary to characterize emissions in the band of interest, assess spectrum-sensing performance, and evaluate interference between heterogeneous systems \cite{Voicu2019, Kuester2021,Danneberg2014,Caromi2018,Young2021}.  One issue complicating such studies is the time and expense required to collect high-quality, real-world recordings of \ac{RF} \ac{I/Q} signals; see \cite{Hale2017a,Nelson2021,Sanders2021} for examples.  

Although early-stage development can be executed with simulated data, it is very challenging to develop accurate models of RF \ac{I/Q} time series for deployed \ac{COTS} communication systems from first principles.  Specifically, in typical measurement scenarios, system parameters and implementation details are at best only partially known, and complex system dynamics are difficult to characterize.  For example, with limited information, it is difficult to accurately model effects due to the analog RF front-end, power control and scheduling dynamics, traffic loading, and wireless propagation \cite{Kuester2021}.  

Considering the difficulties of realistic modeling of \ac{I/Q} signals from ``black-box'' systems, there is a need for flexible, data-driven alternatives that can leverage existing data sets.  Unfortunately, there is currently no established state-of-the-art modeling approach for this problem, and experimental testing often relies on idealistic, synthetic signals, \emph{e.g.}, Refs. \cite{Sanders2017, DOT2018, Young2017}.  One way to characterize a data set with unknown properties is through the use of unsupervised deep generative modeling.  

Deep generative models utilize a deep neural network to produce samples from a high-dimensional target distribution defined by a training set.  In recent years, deep generative models, and most notably, \acp{GAN}, have drawn a great deal of attention in the machine-learning community and have progressed rapidly; \emph{e.g.}, see Refs. \cite{Creswell2018, Hong2019, Wang2019, Pan2019, Bond2021} for reviews.  Namely, \acp{GAN} and other deep generative models have been used to successfully synthesize and process realistic images, speech, and video.  Because deep generative models can be trained in an unsupervised manner, they are a potentially flexible and powerful approach for data-driven modeling of \ac{RF} \ac{I/Q} time series with unknown, complex attributes.

To date, most work on deep generative models has focused on computer vision applications with images \cite{Creswell2018, Hong2019, Wang2019, Pan2019, Bond2021}, while time series have received less attention.  Nonetheless, there have been several papers proposing generative models for time series, primarily for audio applications.  Examples of non-GAN deep generative models for audio time series include WaveNet \cite{van_den_oord2016}, an autoregressive model, and MusicVAE \cite{Roberts2018}, which uses a \ac{VAE}. 

Several time-series \ac{GAN} models leverage prior work on \acp{GAN} for images by training the generator to produce an image-domain time-frequency representation, such as a spectrogram, which is then mapped into a time series.  Examples of models employing this approach include SpecGAN \cite{Donahue2019}, MelGAN \cite{Kumar2019}, TSGAN \cite{Smith2020}, GANSynth \cite{Engel2019},  and TiFGAN \cite{Marafioti2019}.  Additional approaches were compared by  Nistal \emph{et al.} \cite{Nistal2021}.  

Alternatively, there have been some papers proposing \acp{GAN} that directly model time series.  One class of methods includes architectures based on \acp{RNN}, such as \ac{LSTM}, \emph{e.g.}, Refs. \cite{Esteban2017, Yoon2019}.  Methods based on \acp{CNN} include WaveGAN \cite{Donahue2019}, which employs a flattened version of the popular DCGAN model  \cite{Radford2015}, and QuantGAN \cite{Wiese2020}, which uses \acp{TCN} \cite{Bai2018}.

In the field of wireless communications, there have been some recent efforts to apply \acp{GAN}.  Examples include \acp{GAN} for data augmentation for signal classification \cite{Davaslioglu2018, Patel2020}, wireless channel modeling \cite{Ye2020}, anomaly detection \cite{Zhou2021}, and adversarial attacks on communication systems \cite{Shi2021, Sagduyu2021}.  Due to their focus on use case performance, \emph{e.g.}, signal classification accuracy, none of the aforementioned works presents comprehensive evaluations of generated signal fidelity, nor do they compare their models to other approaches.  Furthermore, these prior works did not investigate how their models performed with respect to increasing data set complexity and signal length.  As such, they provide little insight into generative model effectiveness and associated performance limitations.

In this paper, we present the first in-depth investigation of generated signal fidelity for \acp{GAN} applied to unsupervised modeling of baseband \ac{I/Q} \ac{OFDM} signals, where each subcarrier is digitally modulated with \ac{QAM} \cite{Proakis2002, Molisch2012}; see Sec.~\ref{sec:comms_background} for background.  Because OFDM is a commonly used modulation and encoding scheme for digital transmission that is used in cellular networks and wireless local area networks (WLANs) \cite{3GPP-PHY,IEEE-WLAN}, investigations with OFDM signals provide a strong basis for generative modeling of communication signals.  Using synthetic data with known ground truth, we investigated model performance over a range of OFDM parameters and conditions, including fading channels.  

The findings presented here with synthetic data are anticipated to provide a foundation for future investigations into generative models trained with real-world recordings of RF time series and other types of communications signals.  In particular, we are primarily interested in using GANs to generate waveforms for interference testing as well as for data obfuscation.  By focusing on experiments with simulated data, we are able to gain important early-stage insights into model effectiveness that would not otherwise be possible. Namely, the ease of creating unlimited amounts of synthetic data and the ability to control OFDM parameters allow us to investigate the ability of models to handle increasing data set complexity.  Moreover, because synthetic data enable straightforward symbol demodulation, it is possible to directly evaluate QAM symbol constellation fidelity.  These findings indicate which use cases that are feasible for unsupervised generative models as well as those requiring additional advances.\footnote{Python code implementing our models and experiments is available at \url{https://github.com/usnistgov/OFDM-GAN}.  Also, experimental results are available at \url{https://doi.org/10.18434/mds2-2532}.} 

\section{Background: Digital Communication Signals}
\label{sec:comms_background}

We review relevant aspects of digital modulation and wireless channel models in this section.  Further background can be found in textbooks on wireless communications, \emph{e.g.}, Refs. \cite{Proakis2002, Molisch2012}.

\subsection{Baseband Signal Representation}
For a typical narrowband wireless communication system, where the signal bandwidth is much smaller than the carrier frequency, $f_c$, the transmitted bandpass signal can be expressed in the alternative forms

\begin{subequations}
\begin{align}
    x(t) &=  r(t)\cos(2\pi f_c t + \phi(t)) \\
        &= Re\{s(t)\exp(j2\pi f_c t)\} \\
        &= Re\{[I(t) + jQ(t)]\exp(j2\pi f_c t)] \} \\
        &= I(t)\cos(2\pi f_c t) - Q(t)\sin(2\pi f_c t).
\end{align}
\end{subequations}

Above, $r(t)\geq 0$ and $\phi(t)$ are the time-varying amplitude and phase, respectively, $j^2=-1$, and $Re\{\cdot\}$ denotes the real part of a complex number.  The quantity, $s(t)=r(t)\exp[j\phi(t)]$, called the baseband or lowpass representation of the signal, can be decomposed as $s(t) = I(t) + jQ(t)$, where the real and imaginary components are called the in-phase and quadrature (I/Q) components of the signal, respectively \cite{Proakis2002}.

\subsection{Quadrature Amplitude Modulation}
Digital modulation maps a sequence of data bits into a physical waveform.  A commonly used digital modulation method, called quadrature amplitude modulation (QAM), can be interpreted as a combination of digital-amplitude and digital-phase modulation \cite{Proakis2002}.  Namely, for M-ary \ac{QAM} (M-QAM), the transmitted bandpass signal for the $m$th symbol takes the form 
\begin{subequations}
\begin{align}
    x_m(t) &= A_m g_T(t) \cos(2\pi f_c t) + B_m g_T(t) \sin(2\pi f_c t) \\
    &= Re\{C_m g_T(t) \exp(j 2\pi f_c t)\}
\end{align}
\end{subequations}
for $m=1, 2, \ldots, M$, where $A_m$ and $B_m$ are real-valued amplitude levels, $C_m  = A_m - jB_m$ is the corresponding complex-valued amplitude, and $g_T(t)$ is a real-valued pulse-shape function of duration $T$ \cite{Proakis2002}.  The set of complex amplitudes, $C_m$, can be visualized as points in an I/Q-space constellation diagram, which is typically a rectangular grid with $M$ equal to a power of two.  In this case, each symbol represents $k=\log_2(M)$ bits.  Gray encoding, in which adjacent symbols differ by only one bit, is a preferred method to map $k$ bits to a symbol \cite{Proakis2002}.

\subsection{Orthogonal Frequency-Division Multiplexing}
\label{sec:OFDM_background}

Modern wireless communication systems aim to achieve high data rates across wireless channels with time-varying, frequency-selective distortions.  One digital modulation scheme that was designed for such conditions is orthogonal frequency-division multiplexing (OFDM) \cite{Molisch2012,Proakis2002}.  \ac{OFDM} converts a high-rate data sequence into multiple low-rate data sequences that are transmitted over parallel, narrowband channels that can be easily corrected (equalized) for channel distortions \cite{Molisch2012}.  \ac{OFDM} is used by many modern communication systems, including cellular networks \cite{3GPP-PHY} and \acp{WLAN} \cite{IEEE-WLAN}.  

\ac{OFDM} modulates symbols in parallel across $N_{sc}$ subcarriers.  Specifically, a discrete time, baseband \ac{OFDM} signal for a single symbol period, $T_s$, has the (unnormalized) form
\begin{equation}
    s(t_k) = \sum_{n=0}^{N_{sc}-1} C_n \exp\left(j2\pi n k/N\right),
\label{eq:OFDM}
\end{equation}
where $t_k = kT_s/N_{sc}$ (for $k=1, 2, \ldots, N_{sc}$) are discrete time samples, and $C_n$ is the complex-valued symbol for the nth subcarrier \cite{Proakis2002,Molisch2012}.  When the symbols are drawn from a QAM constellation, the resulting scheme is called \ac{OFDM}/QAM, where classic \ac{OFDM}/QAM uses a rectangular pulse shape \cite{Du2007}.  Because Eq. (\ref{eq:OFDM}) is simply an inverse \ac{DFT} on the block of $N_{sc}$ transmit symbols, \ac{OFDM} is usually implemented with the inverse \ac{FFT} algorithm \cite{Proakis2002,Molisch2012}.

Multipath propagation in wireless channels causes delay dispersion and frequency-selective fading \cite{Molisch2012}.  Delay dispersion gives rise to \ac{ISI}.  Also, in the case of multiple-carrier systems, such as \ac{OFDM}, delay dispersion leads to a loss of orthogonality between subcarriers, resulting in \ac{ICI} \cite[Sec. 19.7]{Molisch2012} \cite{Chen2004}.  Standard implementations of \ac{OFDM}/QAM alleviate both \ac{ISI} and \ac{ICI} by using a special type of guard interval called a cyclic prefix \cite{Molisch2012,Proakis2002}.  A cyclic prefix is a copy of the last part of the symbol of duration $T_{cp}$ that is prepended to the symbol, so that the total symbol duration is $T_{cp} + T_s$.  At the receiver, the cyclic prefix, which may be corrupted by delay dispersion, is discarded.  Assuming that the channel is static for the duration of an \ac{OFDM} symbol and that the cyclic prefix is long enough, the effects of delay dispersion are alleviated, reducing \ac{ISI} and \ac{ICI}.  The resulting \ac{OFDM} system can be modeled by a set of parallel nondispersive, fading channels, each with its own complex-valued transfer coefficient.  Consequently, equalization (correction) of the \ac{OFDM} symbol for channel effects is very simple and can be carried out with element-wise division by the transfer function in the frequency (subcarrier) domain.  The frequency-domain transfer function is typically estimated by using pilot symbols \cite{Molisch2012}.    

\subsection{Multipath Channel Models}
\label{sec:background_channel_models}

Multipath propagation in wireless channels results in frequency-selective fading; \emph{i.e.}, the transfer function of the channel varies in frequency.  A widely used class of statistical models for multipath propagation relies on the \ac{WSSUS} assumption \cite{Molisch2012, Salous2013}.  A \ac{WSSUS} channel can be represented as a tapped-delay-line filter, where the coefficients of each tap vary in time.  Namely, the channel impulse response is
\begin{equation}
    h(t,\tau) = \sum_{i=1}^{N} c_i(t)\delta(\tau-\tau_i),
\end{equation}
where $N$ is the number of multipath components, $c_i(t)$ are time-dependent coefficients, $\delta(t)$ is the Dirac delta function, and $\tau_i$ is the time delay of the ith tap \cite{Molisch2012}.  The most commonly used version of this model, the $N$-tap Rayleigh fading model, takes the coefficients, $c_i(t)$, to be zero-mean complex Gaussian random processes with autocorrelation functions determined by associated Doppler spectra, \emph{e.g.}, the classical Doppler spectrum \cite{Molisch2012}.    
A well-known collection of $N$-tap Rayleigh fading models is specified in the \ac{3GPP} cellular standard \cite[Annex B.2]{3GPP-channels}.  We used these models to simulate the effects of multipath propagation in the fading channel experiment presented in Sec.~\ref{sec:channel_experiment}.

\section{Background: Generative Adversarial Networks}

Here, we summarize key features of \ac{GAN} models and Wasserstein loss with gradient penalty, the loss function that we used for training our models.  Familiarity with the fundamentals of deep neural networks is assumed.  For general background on deep learning, see the textbook by Goodfellow \emph{et al.} \cite{Goodfellow2016}.

\subsection{Generative Adversarial Networks}

A \ac{GAN} consists of two neural networks, a generator, $G$, and a discriminator, $D$.  The generator is trained to generate samples from a target data distribution, and the discriminator attempts to distinguish between generated and target data samples.  Specifically, the generator, $G(z)$, learns to map latent vectors, $z$, drawn from a multidimensional Gaussian distribution, $p_Z(z)$, to the generator distribution, $p_g$.  The discriminator maps sample data, $x$, to $D(x)$, representing the probability that a sample belongs to the target distribution, $p_d$.  

In the original \ac{GAN} formulation \cite{Goodfellow2014}, the generator and discriminator compete against each other in the form of a zero-sum game, although it was soon discovered that the original GAN model can suffer from training instabilities, such as diverging loss and mode collapse.  Subsequently, a large number of papers were published that attempted to address these shortcomings; see Refs. \cite{Creswell2018, Hong2019, Wang2019, Pan2019, Bond2021} for reviews.  

\subsection{Wasserstein Loss with Gradient Penalty}
\label{sec:WGAN-GP}

A popular method that has been found to help stabilize \ac{GAN} training is the Wasserstein loss function used by Arjovsky \emph{et al.} \cite{Arjovsky2017}.  The \ac{WGAN} aims to minimize the Wasserstein distance between the generated distribution and the target data distribution, where the Wasserstein ``earth-mover's'' distance can be interpreted as the minimum cost of transporting mass to transform one distribution into another.  Due to the intractability of computing the Wasserstein distance directly, Arjovsky \emph{et al.} instead proposed to train the generator to minimize the proxy loss
\begin{equation}
        L_W = \max_{D \in \mathcal{D}}~E_{x\sim p_{d}}[D(x)] - E_{z\sim p_Z}[D(G(z))],
\label{eq:Wasserstein_loss}
\end{equation}
where $\mathcal{D}$ is the set of 1-Lipshitz functions, and $E[\cdot]$ denotes expected value.  When the discriminator is trained to optimality, \emph{i.e.}, the maximum is attained above, then minimizing the value function in Eq. (\ref{eq:Wasserstein_loss}) with respect to the generator parameters, \emph{i.e.}, the neural network weights, is equivalent to minimizing the Wasserstein distance between $p_g$ and $p_d$ \cite{Arjovsky2017, Gulrajani2017}.  Following common practice, the loss function is minimized using the widely used back-propagation algorithm \cite[Sec.~6.5]{Goodfellow2016}.  Arjovsky \emph{et al.} enforced the Lipshitz constraint via weight clipping following every training update.

To further stabilize the \ac{WGAN} model, Gulranjani \emph{et al.} \cite{Gulrajani2017} proposed to add a gradient penalty term to the \ac{WGAN} loss function instead of weight clipping.  The resulting loss, denoted \ac{WGAN}-GP, minimizes the objective 
\begin{multline}
        \mathcal{L} = E_{\tilde{x}\sim p_g}[D(\tilde{x})] - E_{x\sim p_{d}}[D(x)] \\+ \lambda~E[(\|\nabla_{\hat{x}}D(\hat{x})\|_2 - 1)^2],
\label{eq:WGP_loss}
\end{multline}
where $\hat{x} = \epsilon x + (1-\epsilon)\tilde{x}$ with $\tilde{x} \sim p_g$, $x \sim p_d$, and $\epsilon \sim U[0,1]$; \emph{i.e.}, $\epsilon$ is drawn from a uniform distribution over the unit interval.  Note that $\hat{x}$ is a random linear interpolation between a real data sample, $x$, and a generated data sample, $\tilde{x}$.

Gulrajani \emph{et al.} \cite{Gulrajani2017} made several implementation recommendations for \ac{WGAN}-GP, which we largely followed; see Sec.~\ref{sec:training}.  First, since the \ac{WGAN}-GP objective penalizes the gradient of the discriminator for each batch independently, the use of batch normalization is not recommended.  Second, like Arjovsky \emph{et al.} \cite{Arjovsky2017}, Gulrajani \emph{et al.} used an imbalanced discriminator-generator update rule, where the discriminator weights were updated five times for each generator update.  Third, they recommended using $\lambda=10$ for the default gradient penalty weight.  Last, Gulrajani \emph{et al.} recommended the \textsc{Adam} optimizer \cite{Kingma2015} for discriminator and generator training with default hyperparameter settings $\alpha=10^{-4}$, $\beta_1 = 0$, and $\beta_2 = 0.9$ for the learning rate and moment decay rates, respectively.    

\section{Synthetic OFDM Data Sets}
\label{sec:datasets}

Later, in Sec.~\ref{sec:experiments}, we present four experiments with synthetic \ac{OFDM} data.  This section describes high-level implementation details for our data simulations.  Specific parameter settings used in each experiment are given in Sec.~\ref{sec:experiments}.   

We used synthetic (simulated) data sets of \ac{OFDM} waveforms, which offer several advantages for early-stage investigations into \ac{GAN} models.  First, since high-quality recordings of real-world OFDM-based communication waveforms are not readily available, and since acquiring such recordings requires significant effort, the ease of creating unlimited amounts of synthetic data is well suited to model development.  Second, synthetic data provide control over multiple OFDM parameters, including OFDM symbol length, cyclic prefix, pilot symbols, and resource allocation size, \emph{i.e.}, the number of occupied subcarriers in an OFDM symbol.  Last, because real-world communication system recordings involve complicated signaling protocols and suffer from nonideal physical effects, implementing software-based channel equalization and demodulation is difficult.  By contrast, synthetic data enable straightforward symbol demodulation and performance evaluation.  

Synthetic data sets are constructed by first simulating a different random sequence of bits for each OFDM waveform, where $0$ and $1$ occur with equal probability.  For $M$-ary QAM, each group of $k=\log_2(M)$ bits is mapped using Gray encoding to QAM symbols \cite{Proakis2002}.  Each block of QAM symbols is then mapped onto a specified collection of OFDM subcarriers, which are modulated into a baseband, \ac{I/Q}, time-domain waveform by applying an inverse \ac{FFT}, producing the multicarrier \ac{OFDM} symbol \cite{Proakis2002}.

Every synthetic waveform consists of a sequence of six \ac{OFDM} symbols, each with a cyclic prefix equal to $25$\,\% of the \ac{OFDM} symbol length.  The \ac{OFDM} symbol length, \emph{i.e.}, the number of subcarriers, is set to be $128$, $256$, or $512$, yielding full time-series lengths of $960$, $1920$, or $3840$, respectively.\footnote{Including the cyclic prefix, each symbol has a length of $160$, $320$, or $640$, respectively.  So for a waveform consisting of six symbols, the time series has a length of $960$, $1920$, or $3840$, respectively.}  The above choices were motivated by the specifications for downlink \ac{LTE} \cite{LTEnutshell,3GPP-BS,3GPP-PHY}.  Namely, symbol lengths of $128$, $256$, and $512$ correspond to \ac{LTE} channel bandwidths of $1.4$\, MHz, $3$\,MHz, and $5$\,MHz, respectively \cite{3GPP-BS}.  The cyclic prefix size corresponds to the so-called ``extended'' cyclic prefix option in LTE, for which there are six OFDM symbols per 0.5\,ms ``slot'' \cite{LTEnutshell}.\footnote{For LTE, the physical sampling rate depends channel bandwidth \cite{LTEnutshell}.}

We considered three different settings for the proportion of occupied OFDM subcarriers, \emph{i.e.}, the resource allocation.  Namely, we set the proportion of occupied subcarriers equal to $25\,\%$, $50\,\%$, or $75\,\%$ of the maximum allowed for downlink LTE \cite[Table~1]{LTEnutshell}.  In each case, the block of occupied subcarriers was centered in frequency, and the zero-frequency (DC) subcarrier was not used.  For OFDM symbol lengths of $128$, $256$, and $512$, the maximum number of occupied subcarriers, excluding the DC subcarrier, was taken to be $75$, $150$, and $300$, respectively \cite[Table~1]{LTEnutshell}.  We refer to the three allocation sizes of $25\,\%$, $50\,\%$, or $75\,\%$ as small, medium, and large allocations, respectively.

To simulate the effect of thermal noise, \ac{AWGN} was added to the \ac{OFDM} signals.  In each experiment, the \ac{AWGN} level was set such that the \ac{EVM}, as defined in Sec.~\ref{sec:eval_metrics}, was a specified level.  

\begin{figure}[t!]
\centering
\includegraphics[width=\linewidth]{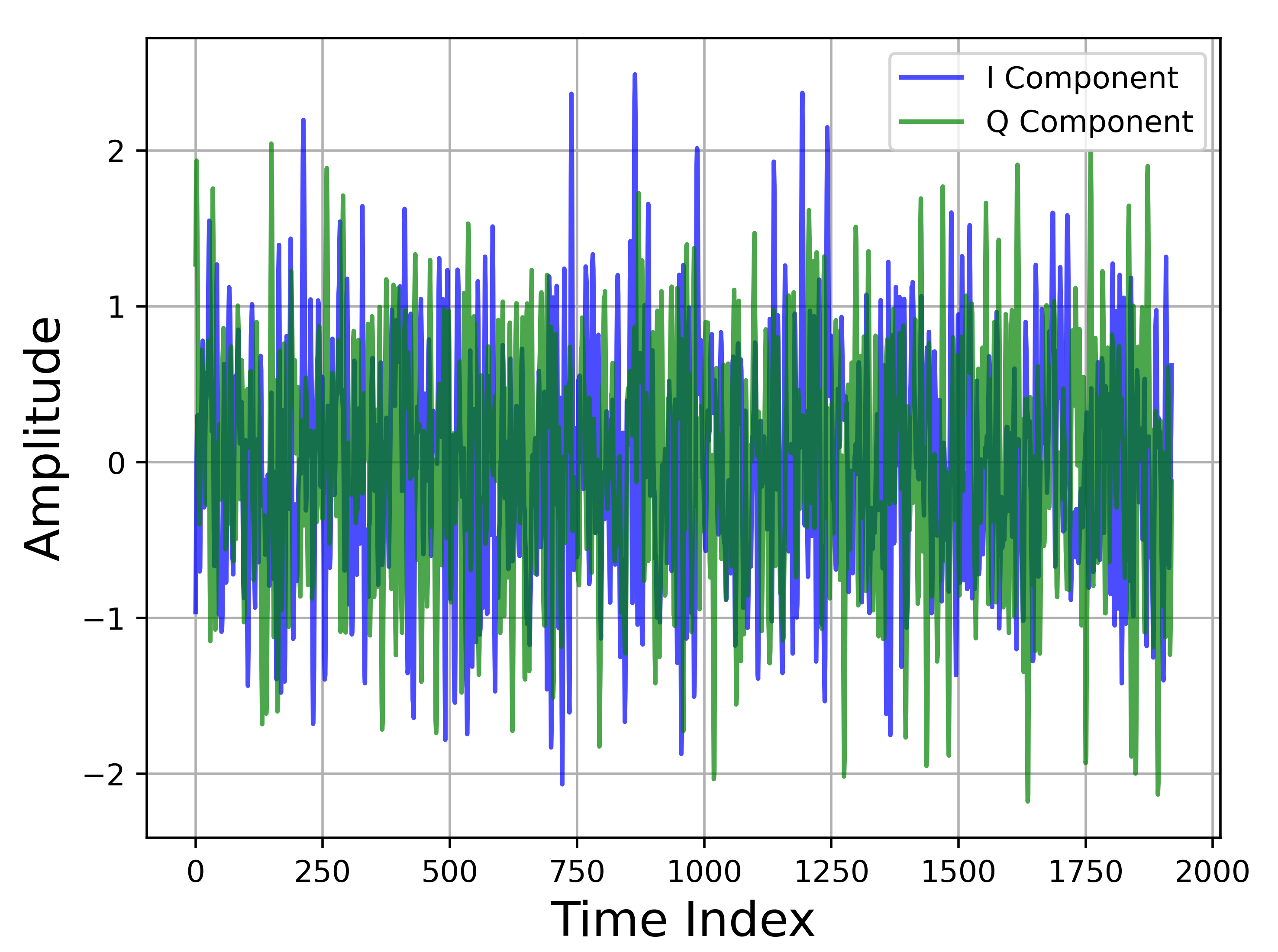} \\
\includegraphics[width=0.49\linewidth]{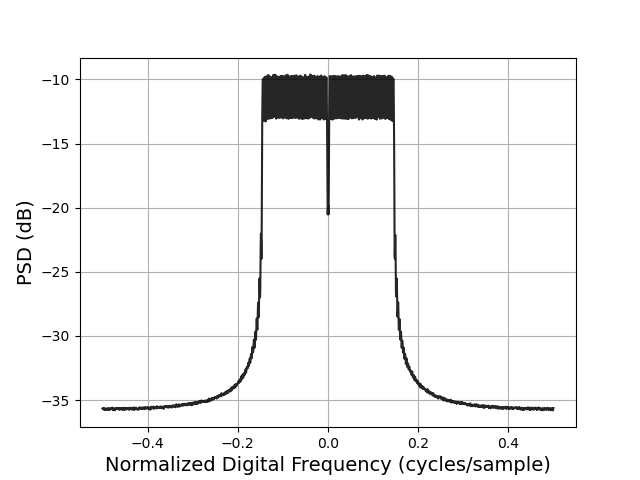}
\includegraphics[width=0.49\linewidth]{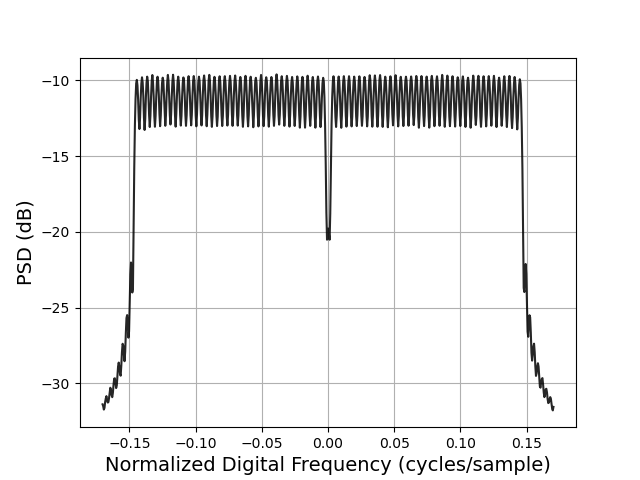}
\caption{Top: Example synthetic OFDM waveform with symbol length of $256$, medium allocation size, and $\text{EVM}=-25$\,dB. Lower Left: Estimate of the corresponding power spectral density (PSD). Lower Right: Zoom of PSD focused on occupied subcarriers.}
\label{fig:target_waveform_example}
\end{figure}

Figure~\ref{fig:target_waveform_example} shows an example synthetic OFDM waveform together with a corresponding estimate of the \ac{PSD}.  Here, the OFDM symbol length is 256, 16-QAM is used on each occupied subcarrier, $\text{EVM} = -25$\,dB, and the allocation size is medium ($50\,\%$ occupancy).  The dip in the \ac{PSD} at zero frequency arises from the fact that the DC subcarrier is not used. 

For our last two experiments in Sec.~\ref{sec:experiments}, we performed additional data transformations to simulate physical effects.  Namely, the third experiment, presented below in Sec.~\ref{sec:channel_experiment}, applied 3GPP fading channel models \cite[Annex B.2]{3GPP-channels} to the synthetic \ac{OFDM} data.  In addition to application of the random channel model, we modified synthetic data generation for this experiment by inserting a pilot symbol on the 4th OFDM symbol to enable estimation of channel frequency response and equalization coefficients \cite{Molisch2012}.  Here, the pilot symbol was taken to be the first Zadoff-Chu base sequence, as defined for the \ac{DMRS} in uplink LTE \cite[Sec.~5.5]{3GPP-PHY}.  Complex-valued Zadoff-Chu sequences are commonly used for channel estimation because they have constant power in time and frequency \cite{Dahlman2013}.  

The fourth experiment, presented in Sec.~\ref{sec:RF_impairment_experiment}, examined two different types of RF impairments that arise from imperfections in analog RF transmitter and receiver hardware.  Namely, we applied models for carrier frequency offset (CFO) and I/Q imbalance; see Smaini \cite{Smaini2012} for background and details.

CFO originates from a frequency offset in the local oscillator between the transmitter and receiver.  CFO in the received baseband signal can be modeled using the relation \cite{Smaini2012}
\begin{equation}
    y(t) = x(t)\exp[j2\pi \Delta_{\text{CFO}} t],
\end{equation}
where $x(t)$ is the ideal baseband signal, $y(t)$ is the CFO-impaired baseband signal, and $\Delta_{\text{CFO}}$ is the CFO spectral shift. To implement this transformation in the discrete-time domain, we set 
\begin{equation}
    y_n = x_n \exp[j2\pi \Delta_{\text{CFO}}n/F_s],
\end{equation}
where $\Delta_{\text{CFO}}$ is specified in Hertz, $F_s$ is the sampling rate specified in Hertz, and $n$ is the time index.  

I/Q imbalance (or mismatch) arises from the fact that quadrature mixers are impaired by gain and phase mismatches.  Letting $x_I(t)$ and $x_Q(t)$ denote the in-phase and quadrature components of the ideal baseband signal, the I/Q imbalance in the transmitter can be modeled as \cite[Eq.~(2.118)]{Smaini2012}
\begin{align}
    y(t) &= \frac{1-(\Delta G/2)}{2}\left[\cos(\Delta \phi/2) + 
    j\sin(\Delta \phi/2)\right] x_I(t) \nonumber  \\ &+ 
    \frac{1+(\Delta G/2)}{2}\left[\sin(\Delta \phi/2) + j\cos(\Delta \phi/2)\right]x_Q(t), 
\end{align}
where $y(t)$ is the impaired, complex-valued baseband signal, and $\Delta_G$ and $\Delta \phi$ are the gain and phase mismatches between the in-phase (I) and quadrature (Q) branches of the circuit.  Here, $\Delta_G = |G_I-G_Q|/G_Q$, where $G_I$ and $G_Q$ are the gains for the I and Q branches, respectively.  We applied the above transformation in the discrete-time domain.  For simplicity, we only considered I/Q imbalance in the transmitter; a similar model can also be used for I/Q mismatch in the receiver \cite{Smaini2012}. 

\section{Novel Generative Model Architectures}
\label{sec:novel_GAN_models}

Here, we present two novel GAN models for OFDM signals that build on prior convolutional GANs for time series.  Specifically, we propose a direct time-series model and another indirect model based on an image-domain time-frequency representation.  Key considerations in our designs included conceptual simplicity and scalability, \emph{i.e.}, the ability of the model to scale to longer and more complex signals, in terms of both computation and performance.  

For many practical applications, the OFDM symbol length is known \textit{a priori}; \emph{e.g.}, it is determined by the channel bandwidth for 4G LTE.  For this reason, the two models presented below are tailored to OFDM data in the sense that they utilize prior knowledge of the OFDM symbol length.  However, all other aspects of the OFDM waveforms, \emph{e.g.}, the QAM symbol constellation, cyclic prefix size, OFDM symbol boundaries, and channel distortions, are assumed to be unknown. 

\subsection{Progressively Scaled Kernel GAN (PSK-GAN)}
\label{sec:PSK-GAN}

Our direct time-series model, called \ac{PSK-GAN}, uses one dimensional (1-D) convolutional layers and aims to model temporal dynamics by progressively scaling kernel lengths with model depth.  Specifically, \ac{PSK-GAN} employs kernels with lengths progressively scaled up or down by a factor equal to the convolution stride with generator and discriminator model depth, respectively.  The motivation behind progressively scaling kernel lengths is to increase the receptive field while avoiding kernels with lengths longer than their inputs.  The process of progressively scaling kernel lengths has the effect of scaling the kernel resolution with feature map resolution.  The concept of progressively scaling kernel sizes with layer depth was motivated by WaveNet \cite{van_den_oord2016}, which employs dilated convolutions to achieve a similar outcome.  

To avoid the generation of so-called ``checkerboard artifacts,'' which manifest as spikes in the power spectrum, convolutional layer kernel lengths are set to be integer multiples of the stride length, as recommended by Odena \emph{et al.} \cite{odena2016deconvolution}.  Based on empirical testing, we set the maximum kernel length equal to the \ac{OFDM} symbol length and the minimum allowable kernel length to $4$. 

Tables~\ref{tab:PSK-GAN-gen} and \ref{tab:PSK-GAN-disc} outline the \ac{PSK-GAN} architectures for the generator and discriminator, respectively.  In these tables, Dense, Conv\,1-D and Transpose Conv\,1-D, denote dense fully connected layers, one-dimensional convolutional layers, and transposed convolutional layers, respectively.  Also, Tanh, ReLU, and LReLU indicate hyperbolic-tangent (Tanh), rectified linear unit (ReLU), and leaky rectified linear unit (LReLU) activation functions.  The filter dimensions for convolutional layers correspond to kernel length, number of input channels, and number of output channels, respectively.  Here, $f = 1$, $2$, or $4$ for OFDM symbol sizes of $128$, $256$, and $512$, respectively, and $n$ is the batch size.  Similarly, the filter dimensions for the dense layers correspond to input length and output length respectively.  To yield time-series lengths that are compatible with convolutional layers with strides of $4$, target \ac{OFDM} signals are zero-padded to the nearest power of $2$. 

\begin{table}[t!]
  \centering
    \caption{PSK-GAN generator architecture [$f$ = 1, 2, 4].}
    \label{tab:PSK-GAN-gen}
    \begin{tabular}{l|c|r} 
      \textbf{Operation} & \textbf{Filter Shape} & \textbf{Output Shape}\\
      \hline
      $z$ $\sim$ Uniform($-$1, 1) &  & ($n$, 100)\\
      Dense & (100, 1024$f$) & ($n$, 1024$f$)\\
      Reshape & & ($n$, 1024, $f$)\\
      ReLU & & ($n$, 1024, $f$)\\
      Transpose Conv1-D (stride=4) & (4$f$, 1024, 512) & ($n$, 512, 4$f$)\\
      ReLU & & ($n$, 512, 4$f$)\\
      Transpose Conv1-D (stride=4) & (4$f$, 512, 256) & ($n$, 256, 16$f$)\\
      ReLU & & ($n$, 256, 16$f$)\\
      Transpose Conv1-D (stride=4) & (8$f$, 256, 128) & ($n$, 128, 64$f$)\\
      ReLU & & ($n$, 128, 64$f$)\\
      Transpose Conv1-D (stride=4) & (32$f$, 128, 64) & ($n$, 64, 256$f$)\\
      ReLU & & ($n$, 64, 256$f$)\\
      Transpose Conv1-D (stride=4) & (128$f$, 64, 2) & ($n$, 2, 1024$f$)\\
      Tanh & & ($n$, 2, 1024$f$)\\
    \end{tabular}
\end{table}

\begin{table}[t!]
  \centering
    \caption{PSK-GAN discriminator architecture [$f$ = 1, 2, 4].}
    \label{tab:PSK-GAN-disc}
    \begin{tabular}{l|c|r} 
      \textbf{Operation} & \textbf{Filter Shape} & \textbf{Output Shape}\\
      \hline
      $x$ $\sim$ $G$($z$) & & ($n$, 2, 1024$f$)\\
      Conv1-D (stride=4) & (128$f$, 2, 64) & ($n$, 64, 256$f$)\\
      LReLU($\alpha=0.2$) & & ($n$, 64, 256$f$)\\
      Conv1-D (stride=4) & (32$f$, 64, 128) & ($n$, 128, 64$f$)\\
      LReLU($\alpha=0.2$) & & ($n$, 128, 64$f$)\\
      Conv1-D (stride=4) & (8$f$, 128, 256) & ($n$, 256, 16$f$)\\
      LReLU($\alpha=0.2$) & & ($n$, 256, 16$f$)\\      
      Conv1-D (stride=4) & (4$f$, 256, 512) & ($n$, 512, 4$f$)\\
      LReLU($\alpha=0.2$) & & ($n$, 512, 4$f$)\\
      Conv1-D (stride=4) & (4$f$, 512, 1024) & ($n$, 1024, $f$)\\
      LReLU($\alpha=0.2$) & & ($n$, 1024, $f$)\\
      Reshape & & ($n$, 1024$f$) \\
      Dense & (1024$f$, 1) & ($n$, 1) \\
    \end{tabular}
\end{table}

\subsection{Short-Time Fourier Transform GAN (STFT-GAN)}

As mentioned in Sec.~\ref{sec:introduction}, many time-series \acp{GAN} train the generator to produce an image-domain time-frequency representation that is then mapped into a time series.  Motivated by these approaches, we propose a two-dimensional ($2$-D) convolutional model, called \ac{STFT-GAN}, that is trained on a complex-valued \ac{STFT} representation of the \ac{OFDM} time series.  Similar \Acp{GAN} based on \ac{STFT} representations have also been used for audio generation, \emph{e.g.}, Refs. \cite{Engel2019, Nistal2021}.  Our model differs from these prior works in two ways.  Namely, our model uses different network architectures and it directly uses the complex-valued STFT without additional processing.  Related approaches that apply additional processing to the STFT, \emph{e.g.}, those in Refs. \cite{Engel2019, Nistal2021}, were not found to be advantageous in preliminary tests with synthetic OFDM data.  

The \ac{STFT} (a.k.a. windowed Fourier transform) is computed by dividing the time series into overlapping segments of equal length, applying a window function, and then calculating the \ac{DFT} on each segment \cite{Smith2011, Mallat2009}.  We used a Hann window and $75$\,\% segment overlap.  The Hann window was used since it is a common default choice, and the amount of overlap was selected to be consistent with methods in Refs. \cite{Engel2019, Nistal2021}.  Under these conditions, the \ac{COLA} constraint is satisfied \cite{Smith2011}, and the \ac{STFT} is invertible; \emph{i.e.}, no information is lost.\footnote{Noninvertible STFTs were not considered, since they do not ensure that samples of the generated distribution have the same dimensionality as samples of the target distribution.} 

\ac{OFDM} target waveforms are first zero-padded to the nearest power of $2$ before conversion to an \ac{STFT} representation.  We set the STFT window length equal to the \ac{OFDM} symbol length, since empirical tests indicated that this choice gives superior results.  The STFT values are rescaled to the range $[-1, 1]$ and shifted such that the zero-frequency component is at the center of each DFT window.

The architecture of \ac{STFT-GAN} is based on the DCGAN architecture \cite{Radford2015}, with modifications made to accommodate the nonsquare shape of the \ac{STFT}.  Specifically, \ac{STFT-GAN} is composed of four $2$-D convolutional layers with $4 \times 4$ kernels for both the generator and discriminator; see Tables~\ref{tab:STFT-GAN-gen} and \ref{tab:STFT-GAN-disc}.   In the tables, the notation is similar to that used for \ac{PSK-GAN}, with Conv\,2-D and Transpose Conv\,2-D indicating two-dimensional convolutional and transposed convolutional layers, $f = 1$, $2$, or $4$ corresponding to waveforms with OFDM symbol lengths of $128$, $256$, and $512$, respectively, and $n$ denoting the batch size. 

\begin{table}[t!]
  \centering
    \caption{\ac{STFT-GAN} generator architecture [$f$ = 1, 2, 4].}
    \label{tab:STFT-GAN-gen}
    \begin{tabular}{l|c|r} 
      \textbf{Operation} & \textbf{Filter Shape} & \textbf{Output Shape}\\
      \hline
      $z$ $\sim$ Uniform(-1, 1) &  & ($n$, 100) \\
      Dense & (100, 16384$f$) & ($n$, 16384$f$) \\
      Reshape & & ($n$, 1024, 8$f$, 2) \\
      ReLU & & ($n$, 1024, 8$f$, 2)\\
      Transpose Conv2-D (stride=2) & (4, 4, 1024, 512) & ($n$, 512, 16$f$, 4)\\
      ReLU & & ($n$, 512, 16$f$, 4)\\
      Transpose Conv2-D (stride=2) & (4, 4, 512, 256) & ($n$, 256, 32$f$, 8)\\
      ReLU & & ($n$, 256, 32$f$, 8)\\
      Transpose Conv2-D (stride=2) & (4, 4, 256, 128) & ($n$, 128, 64$f$, 16)\\
      ReLU & & ($n$, 128, 64$f$, 16)\\
      Transpose Conv2-D (stride=2) & (4, 4, 128, 2) & ($n$, 2, 128$f$, 33)\\
      Tanh & & ($n$, 2, 128$f$, 33)\\
    \end{tabular}
\end{table}

\begin{table}[t!]
  \centering
    \caption{\ac{STFT-GAN} discriminator architecture [$f$ = 1, 2, 4].}
    \label{tab:STFT-GAN-disc}
    \begin{tabular}{l|c|r} 
      \textbf{Operation} & \textbf{Filter Size} & \textbf{Output Shape}\\
      \hline
      $x$ $\sim$ $G$($z$) & & ($n$, 2, 128$f$, 33)\\
      Conv2-D (stride=2) & (4, 4, 2, 128) & ($n$, 128, 64$f$, 16)\\
      LReLU($\alpha=0.2$) & & ($n$, 128, 64$f$, 16)\\
      Conv2-D (stride=2) & (4, 4, 128, 256) & ($n$, 256, 32$f$, 8)\\
      LReLU($\alpha=0.2$) & & ($n$, 256, 32$f$, 8)\\
      Conv2-D (stride=2) & (4, 4, 256, 512) & ($n$, 512, 16$f$, 4)\\
      LReLU($\alpha=0.2$) & & ($n$, 512, 16$f$, 4)\\
      Conv2-D (stride=2) & (4, 4, 512, 1024) & ($n$, 512, 8$f$, 2)\\
      LReLU($\alpha=0.2$) & & ($n$, 512, 8$f$, 2)\\
      Reshape & & ($n$, 16384$f$) \\
      Dense & (16384$f$, 1) & ($n$, 1) \\
    \end{tabular}
\end{table}

\section{Training Protocol}
\label{sec:training}

Both \ac{PSK-GAN} and \ac{STFT-GAN} were trained with WGAN-GP loss described in Sec.~\ref{sec:WGAN-GP}.  Like the WGAN-GP training protocol, \ac{PSK-GAN} and \ac{STFT-GAN} were trained using the \textsc{Adam} optimizer \cite{Kingma2015} for discriminator and generator with hyperparameter settings of $\alpha=10^{-4}$, $\beta_1 = 0$, and $\beta_2 = 0.9$ for the learning rate and moment decay rates, respectively.  In a departure from the WGAN-GP protocol, we used a 1:1 update ratio between the discriminator and generator, modified from the original 5:1 ratio. This choice was found to yield better convergence on our target data sets.  We trained each model with a target data set of size $2^{16} = 65\,536$, for $500$ epochs with a batch size of $128$. 

Following common practice with \acp{GAN}, the training data were scaled to the range $[-1, 1]$, which corresponds to the range of the tanh output activation of the generator \cite{Radford2015}.  Specifically, for \ac{STFT-GAN}, all target distributions were scaled using feature-based min-max scaling, which scales the minimum and maximum values of each time step to $[-1, 1]$.  By contrast, for \ac{PSK-GAN}, the target distribution was scaled with min-max scaling using the global minimum and maximum values.  Global min-max scaling was used for \ac{PSK-GAN} since it was found to yield better results.  All outputs from the generator were rescaled back to the original range using the applicable inverse transformation.

\section{Evaluation Methods}
\label{sec:eval_metrics}

Evaluations of \acp{GAN} often focus on subjective assessments of perceptual quality or quantitative metrics that require a suitable feature space defined by, for example, a pretrained model on a standard data set \cite{Xu2018, Borji2019}.  Since OFDM waveforms are not directly human interpretable (\emph{e.g.}, see Fig.~\ref{fig:target_waveform_example}), it is not possible to assess OFDM signal fidelity on a perceptual basis.  Moreover, there are no standard pretrained classification models for time series, so general-purpose quantitative GAN evaluation measures that require a suitable feature space are not easily applied.  Therefore, we focused our evaluations on OFDM-specific signal attributes.  Namely, we evaluated the quality of the PSD, the \ac{QAM} constellation, and the cyclic prefix.  

All evaluations were conducted with test sets that were $1/4$ the size of the training set.  Test sets had size of $2^{14} = 16\,384$.  Test sets of target waveforms were created independently of the training set, and test sets of generated waveforms were created at the completion of training.  

To avoid parametric assumptions, we estimated the \ac{PSD} by applying the multitaper method \cite{Thomson1982,Percival2020}, a versatile nonparametric approach, to the full duration of each waveform.  The number of frequency bins was taken as the next power of 2 greater than or equal to the waveform length.  To obtain a representative PSD estimate across the test set, we took the median value in each frequency bin; see Fig.~\ref{fig:target_waveform_example} for an example median \ac{PSD} estimate.    

Let the median \acp{PSD} for the target and generated distributions be denoted as $P_t(f_d)$ and $P_g(f_d)$, respectively, where $f_d \in [-0.5, 0.5]$ is normalized digital frequency with units of cycles per sample.  To assess the accuracy of $P_g$ relative to $P_t$, we used the ``geodesic distance'' for power spectra proposed by Georgiou \cite{Georgiou2006, Georgiou2007}, defined as
\begin{multline}
    d_g(P_g,P_t) = \\
    \sqrt{\int_{-0.5}^{0.5}\left( \log \frac{P_g(f_d)}{P_t(f_d)}\right)^2 df_d - \left(\int_{-0.5}^{0.5}\log \frac{P_g(f_d)}{P_t(f_d)} df_d \right)^2}.
\end{multline}
The above quantity can be interpreted as the length of a geodesic connecting points on the manifold of \acp{PSD} \cite{Georgiou2006}.  Notably, this distance does not distinguish \acp{PSD} that differ by a constant, positive, multiplicative factor \cite{Georgiou2007}.  Also, note that the first term is equivalent to a difference of log-transformed power spectra, capturing PSD differences across a large dynamic range.  

Denoting the discrete-valued PSDs as $P_g[k]$ and $P_t[k]$, and approximating the integrals with summations, we obtain the discrete form 
\begin{multline}
    d_g(P_g,P_t) \approx \\
    \sqrt{\sum_k\left( \log \frac{P_g[k]}{P_t[k]}\right)^2 \Delta f_d - \left(\sum_l \log \frac{P_g[l]}{P_t[l]} \Delta f_d \right)^2},
\end{multline}
where the summations are taken over the index for the normalized digital frequency grid with step size $\Delta f_d$.  Since the choice of logarithm above is arbitrary, we chose to implement the above formula with a natural logarithm.  

To quantitatively evaluate the quality of the \ac{QAM} constellation, we used \ac{EVM}, which measures the \ac{RMS} deviation of measured symbols from the ideal signal constellation \cite{Vigilante2017,Mckinley2004}.  Namely, we used the commonly used definition \cite{Vigilante2017}
\begin{equation}
    \text{EVM} = \sqrt{\frac{\frac{1}{N_s}\sum_{i=1}^{N_s} | S_{\text{meas},i} - S_{\text{ideal},i} |^2}{\frac{1}{M}\sum_{i=1}^M | S_{\text{ideal},i} |^2}},
\label{eq:EVM_RMS}
\end{equation}
where $N_s$ is the number of symbols in a random symbol sequence, $M$ is the number of unique symbols in the constellation, $S_{\text{meas},i}$ is the $i$th measured symbol, and $S_{\text{ideal},i}$ is the ideal constellation point for the $i$th symbol.  Above, it is assumed that the number of symbols in the sample, $N_s$, is large enough to ensure that all possible symbols and transitions are observed \cite{Vigilante2017}.  To define added noise levels and to assess signal fidelity, we used the above definition of \ac{EVM} expressed in decibels (dB), \emph{i.e.}, $20\log_{10}\text{EVM}$.  We estimated \ac{EVM} for a single waveform using the set of all \ac{QAM} symbols in the waveform and then found the median \ac{EVM} value across all waveforms in a test set.  

\begin{figure}[t!]
\centering
\includegraphics[width=0.5\linewidth]{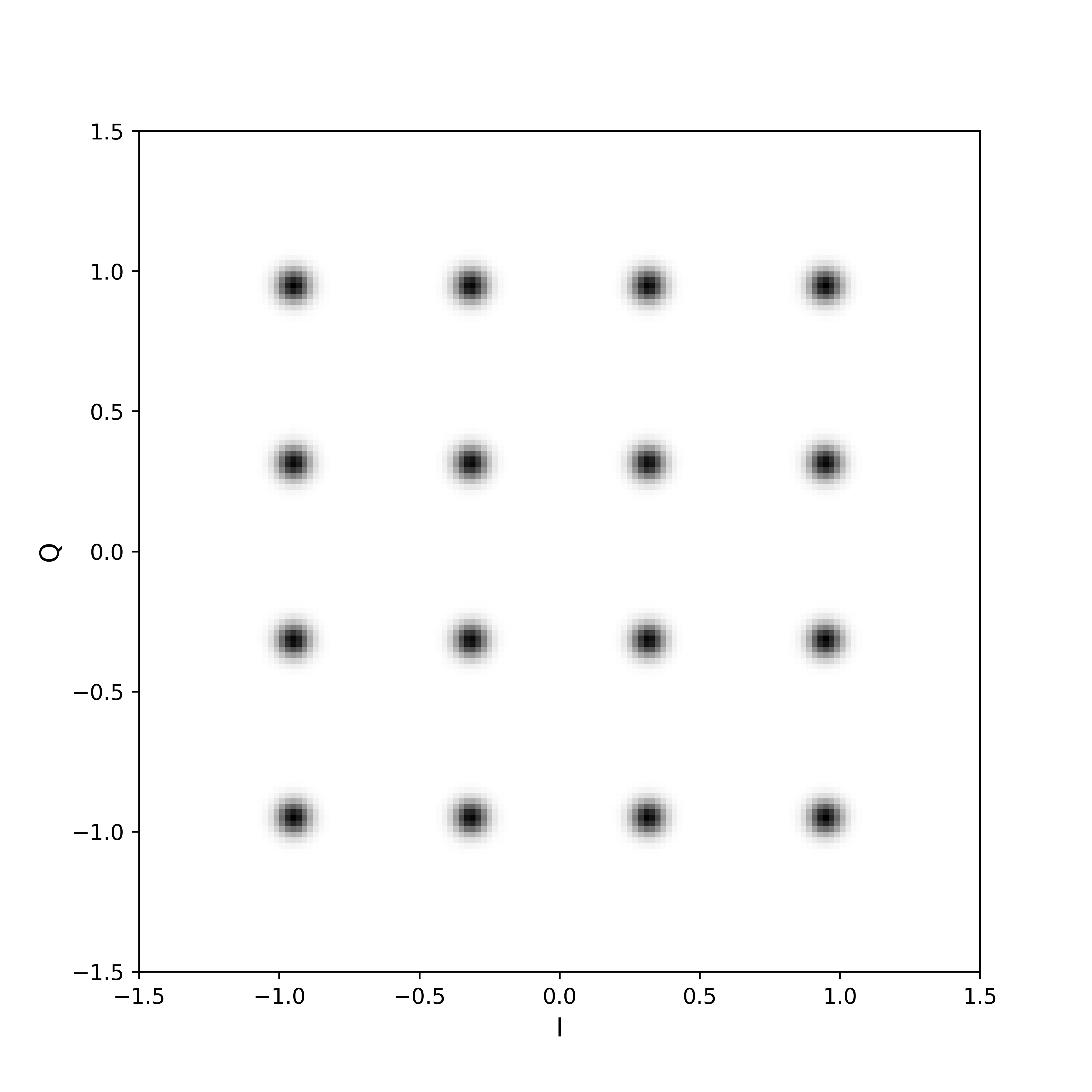}
\caption{Example constellation diagram for a 16-QAM signal constellation with EVM $= -25$\,dB.}
\label{fig:target_constellation}
\end{figure}

We supplemented \ac{EVM} assessment with qualitative evaluation of constellation diagrams, a commonly used method to visualize digitally modulated signals.  A conventional constellation diagram is a scatter plot in I/Q space of the sequence of measured symbols.  Since the total number of QAM symbols in the test set was very large, we plotted 2-D histograms instead of scatter plots.  Specifically, we plotted constellation diagrams using 2-D histograms with $150 \times 150$ bins, evenly spaced over the region $[-1.5, 1.5] \times [-1.5, 1.5]$.  An example 16-QAM constellation diagram for a test set with $\text{EVM} = -25$\,dB is shown in Fig.~\ref{fig:target_constellation}. 

The presence and quality of a cyclic prefix at the beginning of each OFDM symbol were evaluated as follows.  First, for each waveform in the target and generated test sets, we found the cross-correlation function of each cyclic prefix with the waveform, where all cyclic prefix segments were removed.  The location and strength of the cross-correlation maximum indicated the accuracy of the cyclic prefix in the generated waveforms.  We obtained an aggregate metric for cross-correlation strength by finding the median of the maximum cross-correlation values across the generated and target test sets, denoted $\text{R}_{\text{gen}}$ and $\text{R}_{\text{target}}$ respectively, and then computing the relative error, expressed as a percentage, 
\begin{equation}
    \text{RelErr}_{\text{CP}}\% = \frac{| \text{R}_{\text{gen}} - \text{R}_{\text{target}}|}{| \text{R}_{\text{target}}|} \times 100.
\end{equation}

As mentioned earlier in Sec.~\ref{sec:datasets}, our third experiment, presented in Sec.~\ref{sec:channel_experiment}, applied fading channel models to the synthetic OFDM data.  To extract QAM symbols and evaluate the signal constellation, we first equalized the OFDM data on each subcarrier to correct for channel effects, a standard step prior to decoding received OFDM signals.  Namely, the channel frequency response for each block of $6$ OFDM symbols was estimated across occupied subcarriers by taking the demodulated pilot symbol located at the 4th OFDM symbol location and dividing it by the known pilot sequence.  The QAM symbol on each subcarrier was then equalized by dividing by the corresponding estimated channel frequency response coefficient \cite{Molisch2012}.  

A metric commonly used to characterize channels with frequency-selective fading is the coherence bandwidth, defined as the half width at half maximum of the channel's time-frequency correlation function \cite{Molisch2012}.  Because the coherence bandwidth characterizes the correlations between fades at different frequencies, it is particularly relevant for schemes like \ac{OFDM} that transmit on multiple subcarriers \cite{Molisch2012}.  For this reason, we used coherence bandwidth as a performance metric when we investigated fading channels in the experiment in Sec.~\ref{sec:channel_experiment}. 

To estimate coherence bandwidth, we computed the autocorrelation function of the channel frequency response estimated with each pilot symbol and found the half width at half maximum.  We then plotted histograms of the estimated values across the whole test set.  

\section{Experiments}
\label{sec:experiments}

Below, we present the results of four experiments: a data complexity experiment, a modulation-order experiment, a fading channel experiment, and an RF impairment experiment.  

\subsection{Data Complexity Experiment}
\label{sec:main_experiment}

The objective of the data complexity experiment was to evaluate how well our GAN models performed as the target OFDM data set became increasingly complex.  Namely, we changed two experimental factors: OFDM symbol length and the resource allocation size (proportion of occupied subcarriers).  For details on how settings for these factors are implemented, see Sec.~\ref{sec:datasets}.  We used three settings for the OFDM symbol length, 128, 256, and 512, and three settings for the allocation size, denoted small, medium, and large, resulting in a total of nine test configurations.  All target data sets used 16-QAM digital modulation on the occupied OFDM subcarriers, and \ac{AWGN} was added such that $\text{EVM}=-25$\,dB.  This \ac{EVM} value was selected because it corresponds to a noticeable level of noise that results in essentially no bit errors, \emph{i.e.}, a strong communication link \cite{Shafik2006}.  We compared the two GAN models presented in Sec.~\ref{sec:novel_GAN_models}, \ac{PSK-GAN} and \ac{STFT-GAN}, to an implementation of WaveGAN \cite{Donahue2019}, described in the Appendix.  WaveGAN was chosen as a baseline model for comparison because it is a state-of-the art direct time-series GAN. 

To assess training variability, all models were trained on each data set three times, with different neural network weight initialization and different batch randomization.  On our computational hardware, the training time for each model was approximately 12 h.  Therefore, due to the high computational cost associated with training multiple model instances, it was necessary to limit the number of repetitions.  For this reason, three repetitions of each configuration was selected to gain limited insights into training variability.

\begin{figure}[t!]
\centering
\includegraphics[width=0.75\linewidth]{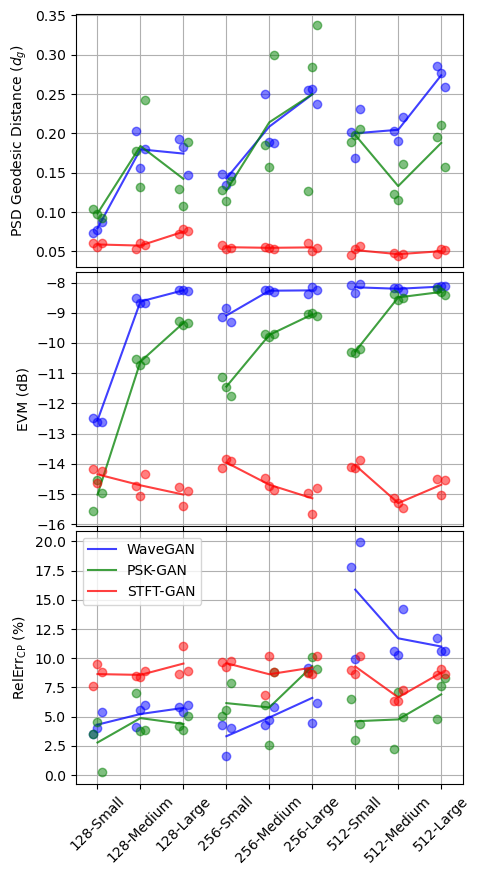}
\caption{Aggregate results for the data complexity experiment.  Top: Power spectral density (PSD) geodesic distance. Middle: Error vector magnitude (EVM). Bottom: Relative error in maximum cyclic prefix cross-correlation.}
\label{fig:agg_results_main_exp}
\end{figure}

Figure~\ref{fig:agg_results_main_exp} summarizes the results of the data complexity experiment for the PSD, EVM, and cyclic prefix evaluations, respectively.  The $x$-axis tick labels indicate the OFDM symbol length and the allocation size; \emph{e.g.}, ``128-Small'' denotes the test configuration with a 128 symbol length and small allocation size.  In these plots, the results for each model repetition are shown with circles.  Also, lines connecting average values across the three repetitions are shown to aid visual interpretation.  Error bars were omitted from the plots since the dominant source of uncertainty was training variability as reflected by the spread of the three repetitions, and the uncertainties for individual model results were too small to be visible.  

The PSD results in Fig.~\ref{fig:agg_results_main_exp} (top) show that \ac{STFT-GAN} consistently had the smallest PSD distance relative to the target distribution and was fairly consistent across test conditions.  On the other hand, \ac{PSK-GAN} and WaveGAN displayed a wider range of performance, with much larger PSD distances in many cases.  

\begin{figure*}[t!]
\centering
\includegraphics[width=0.32\linewidth]{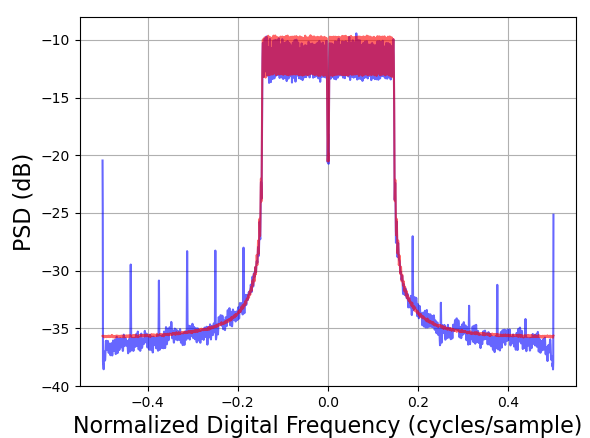}
\hspace{.1cm}
\includegraphics[width=0.32\linewidth]{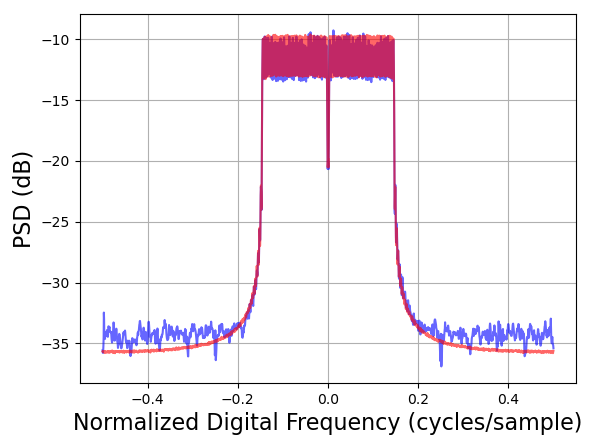}
\hspace{.1cm}
\includegraphics[width=0.32\linewidth]{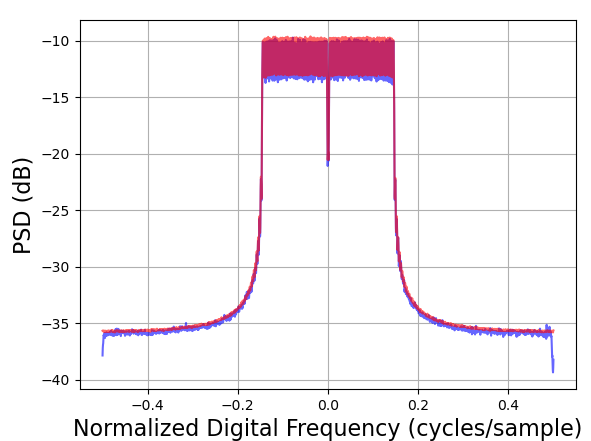}
\caption{Estimated median power spectral densities (PSDs) for the data complexity experiment with 256 OFDM symbol length and medium allocation size.  The PSD for the target distribution is shown in red, and the PSD for the generated distribution is shown in blue.  Left: WaveGAN. Middle: PSK-GAN. Right: STFT-GAN.}
\label{fig:PSD_examples_main_exp}
\end{figure*}

Estimated median PSDs for the 256-Medium condition are shown in Fig.~\ref{fig:PSD_examples_main_exp}.  These plots show that the PSD for the generated distribution from WaveGAN suffered from spikes, likely due to the so-called ``checkerboard artifact'' phenomenon; see the Appendix.  Also, the spectral density for \ac{PSK-GAN} is seen to be higher than the target distribution in areas without occupied subcarriers.  On the other hand, the PSD for \ac{STFT-GAN} shows excellent agreement with the target distribution.   

\begin{figure*}[t!]
\centering
\includegraphics[width=0.32\linewidth]{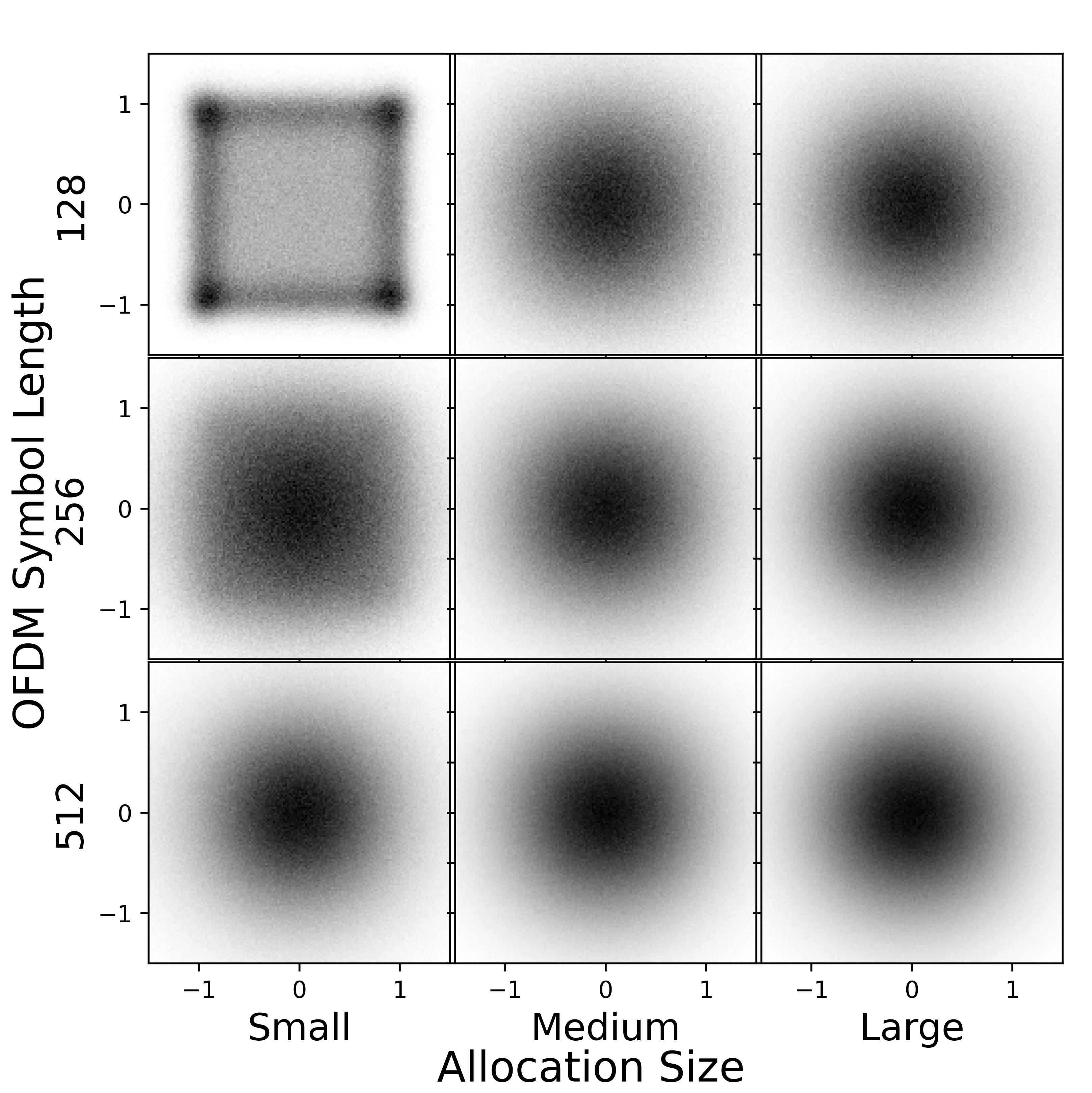}
\hspace{.1cm}
\includegraphics[width=0.32\linewidth]{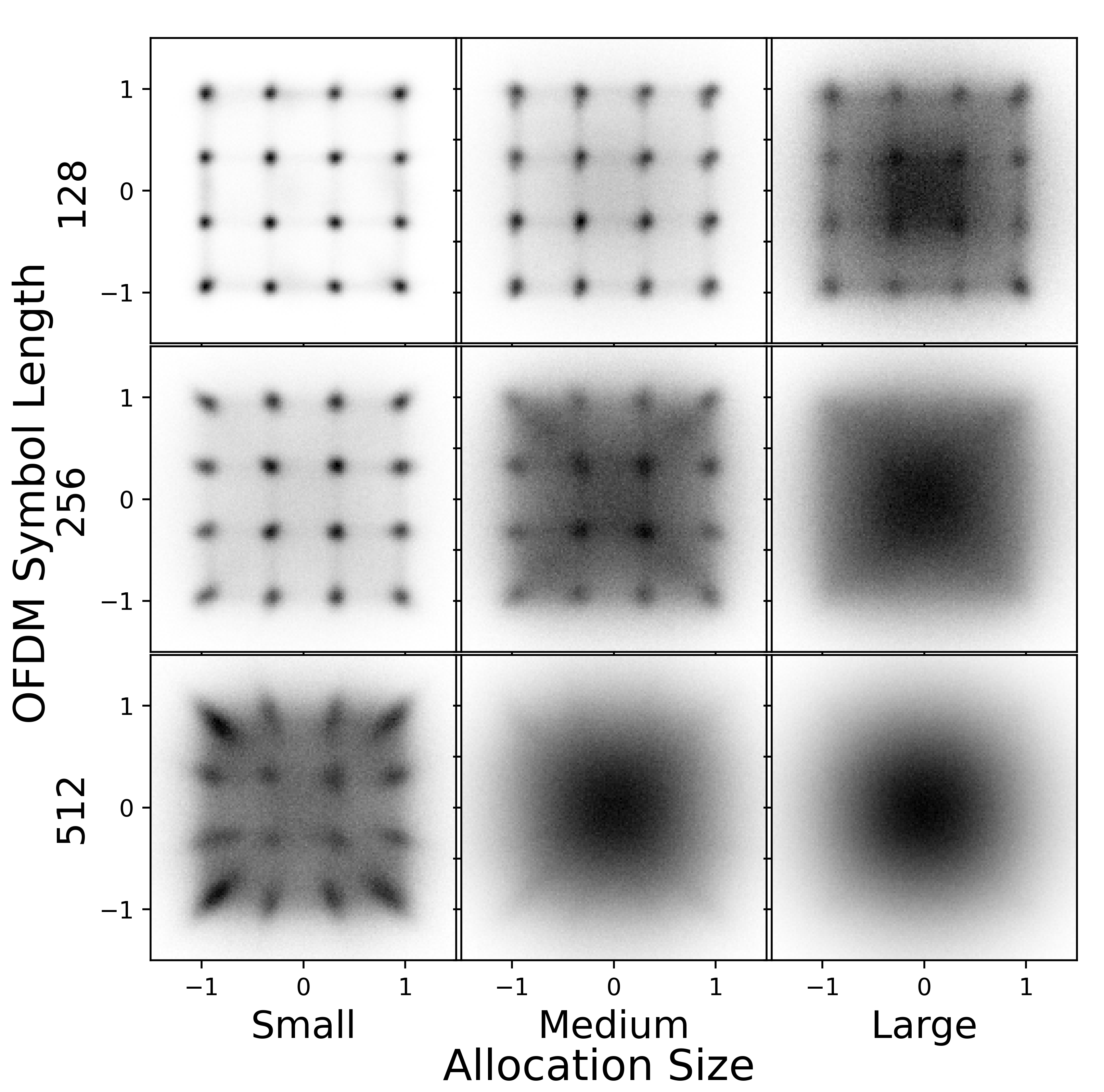}
\hspace{.1cm}
\includegraphics[width=0.32\linewidth]{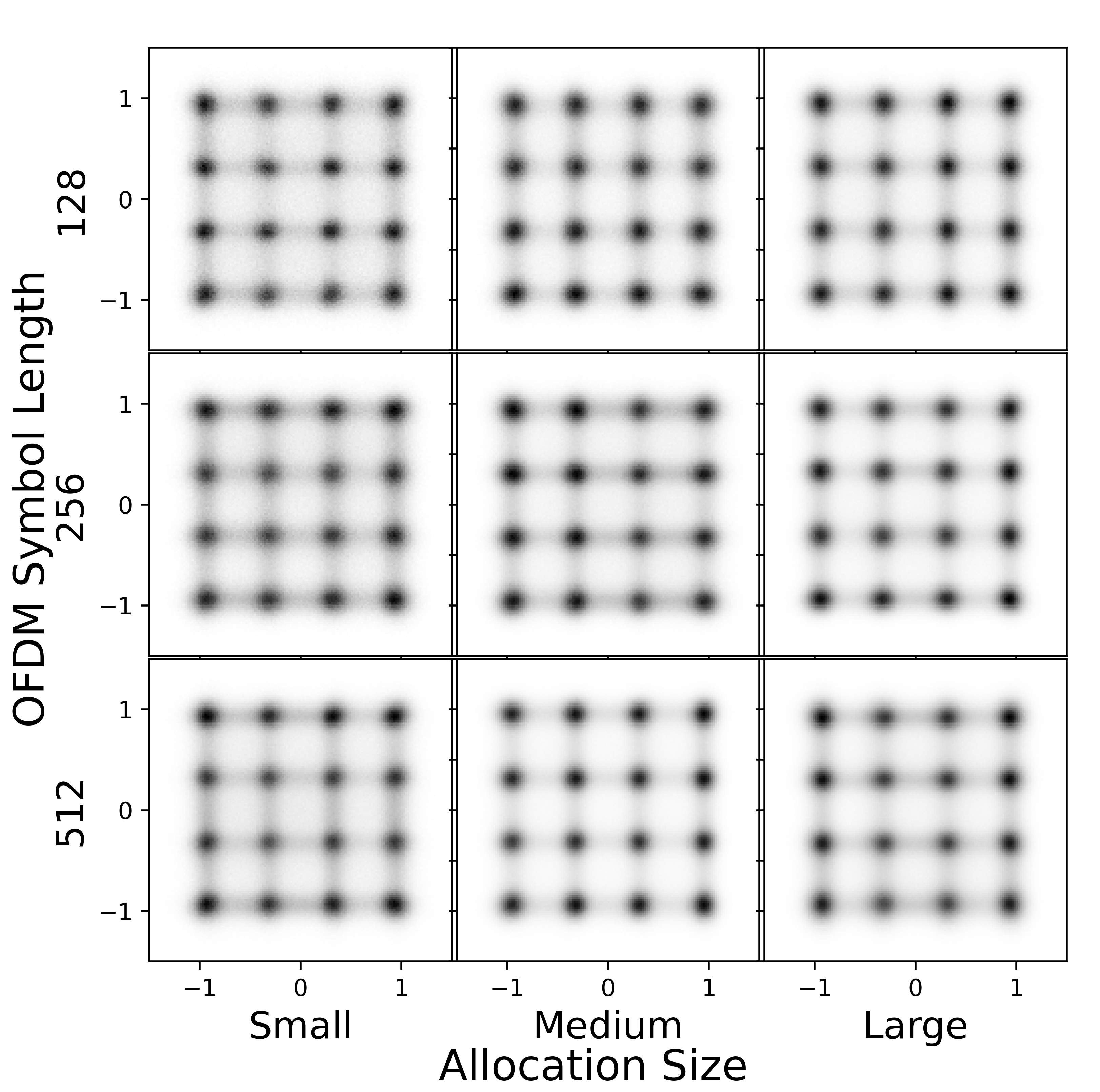}
\caption{Constellation diagrams for the data complexity experiment. Left: WaveGAN. Middle: PSK-GAN. Right: STFT-GAN.}
\label{fig:main_exp_constellations}
\end{figure*}

\begin{figure}[t!]
\centering
\includegraphics[width=0.85\linewidth]{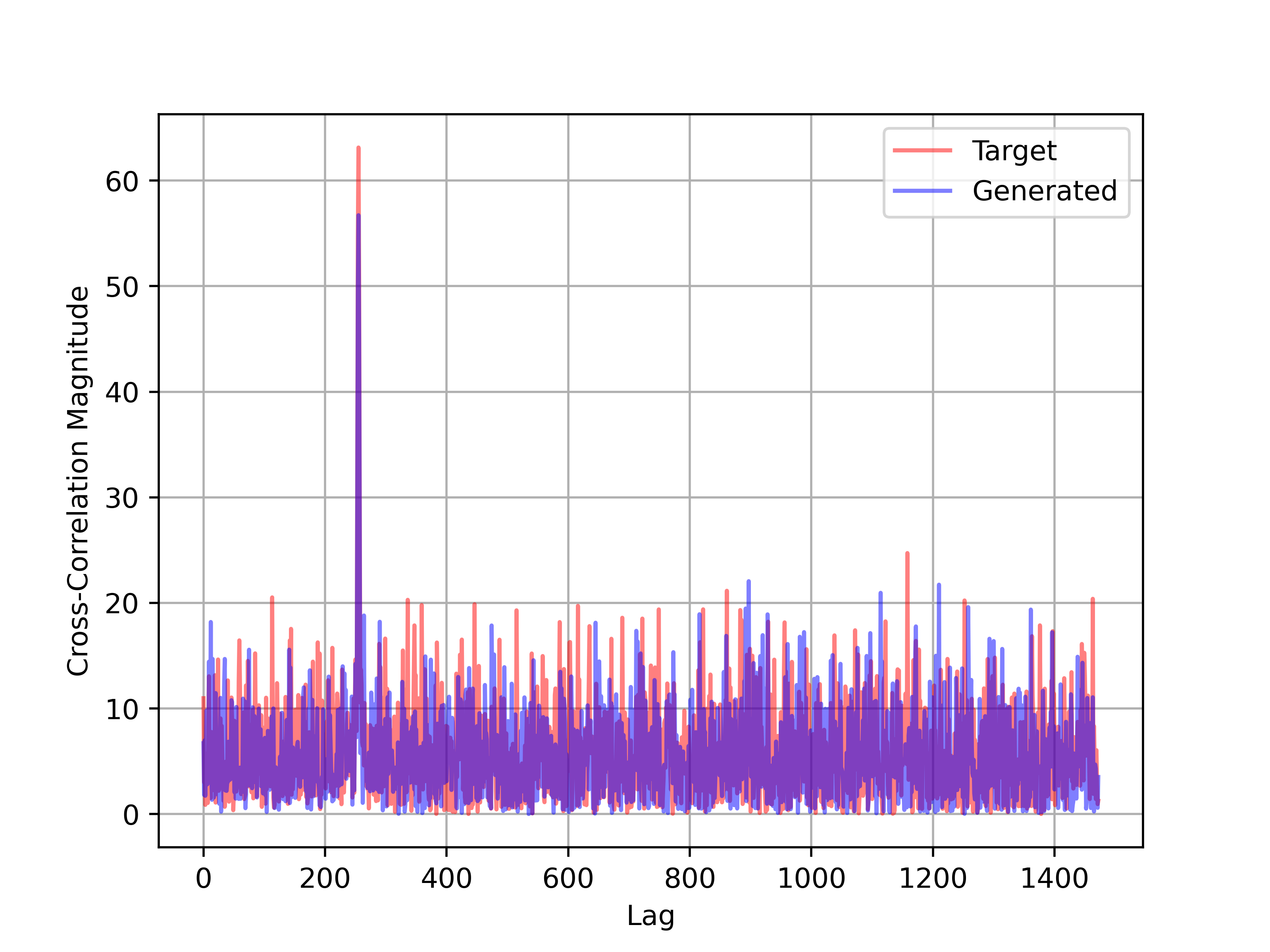}
\caption{Median cross-correlation magnitude of all cyclic prefixes and OFDM symbols in the 256-medium condition for STFT-GAN.}
\label{fig:main_exp_cyclic_prefix}
\end{figure}

The \ac{EVM} results in Fig.~\ref{fig:agg_results_main_exp} (middle) display clear trends in performance.  Specifically, the direct time-series models, \ac{PSK-GAN} and WaveGAN, showed worsening performance as both the OFDM symbol size and allocation size increased, with \ac{PSK-GAN} edging out WaveGAN, especially for small allocation sizes.  By contrast, \ac{STFT-GAN} markedly outperformed the direct time-series models and did not display a degradation in performance as the symbol size and allocation size increased.  The \ac{EVM} achieved by \ac{STFT-GAN} ranged between $-$14\,dB and $-$16\,dB, indicating that the recovered constellation fidelity was worse than the target \ac{EVM} of $-$25\,dB by as much as 9\,dB.  Clues to the source of the lower fidelity are visible in the constellation diagrams described next.

Figure~\ref{fig:main_exp_constellations} displays constellation diagrams for the three GAN models; compare these diagrams to Fig.~\ref{fig:target_constellation} for the target distribution.  The plots in Fig.~\ref{fig:main_exp_constellations} show clear qualitative differences in performance between the models.  Specifically, for all conditions, WaveGAN struggled to learn the full 16-QAM constellation.  \ac{PSK-GAN} successfully learned the 16-QAM constellation in the simpler conditions with smaller allocations and shorter symbol lengths, but it performed worse as the symbol length and allocation size increased.  On the other hand, \ac{STFT-GAN} clearly learned the 16-QAM symbol constellation under all conditions.  However, as noted above for the \ac{EVM} results, the observed constellation fidelity for \ac{STFT-GAN} did not match the target distribution.  For example, there are visible grid lines between symbols that can be interpreted as a form of mode mixing.  

The cyclic prefix results are shown in Fig.~\ref{fig:agg_results_main_exp} (bottom).  WaveGAN performed fairly well for the 128 and 256 symbol lengths, but performance was much worse for the 512 symbol length, with relative errors above 10\,\%.  On the other hand, \ac{PSK-GAN} and \ac{STFT-GAN} had relatively uniform performance across all conditions, with \ac{PSK-GAN} displaying slightly better relative errors below 10\,\%.         

Figure~\ref{fig:main_exp_cyclic_prefix} shows the median cross-correlation of all cyclic prefixes and OFDM symbols for generated and target distributions, respectively, in the 256-medium condition for STFT-GAN.  Here, each cyclic prefix is cross-correlated with the whole waveform with the cyclic prefixes removed.  The lags are normalized so that zero lag corresponds to the first time step of the OFDM symbol associated with each cyclic prefix.  The plot for the generated waveform distribution shows a peak correlation in the expected location at the end of each OFDM symbol.  Also, the cross-correlation plot indicates insignificant correlation with the remaining OFDM symbols.  These results are representative of the findings for all other conditions and models; \emph{i.e.}, all models learned the cyclic prefix location accurately.

Overall, based on the superior performance of \ac{STFT-GAN} as evidenced by the PSD and \ac{EVM} metrics, as well as the constellation diagrams, we conclude that \ac{STFT-GAN} is the most effective of the three models at adapting to increasing OFDM data set complexity.  Note that all three models had a similar number of weights and, hence, comparable computational requirements.

We speculate that the better performance of \ac{STFT-GAN} may be due to the fact that the STFT representation was calculated with an STFT window size equal to the OFDM symbol length, which implies that each pixel resolves a single subcarrier.  On the other hand, the direct waveform models use a data representation in which all subcarriers are superimposed in time.  Due to the superiority of STFT-GAN, the following experiments focused specifically on STFT-GAN.  

\subsection{Modulation-Order Experiment}
\label{sec:mod_order_experiment}

The goal of the modulation-order experiment was to assess how \ac{STFT-GAN} performed with different QAM modulation orders, which are dynamically adjusted in LTE and \acp{WLAN} to adapt the data transmission rate to the propagation channel \cite{3GPP-PHY,IEEE-WLAN}.  We considered modulation orders of $M=4$, $16$, $32$, and $64$, respectively.  In all cases, the OFDM symbol length was set to $128$, the allocation size was medium, and the target \ac{EVM} was $-$25\,dB.  

\begin{figure}[t!]
\centering
\includegraphics[width=.85\linewidth]{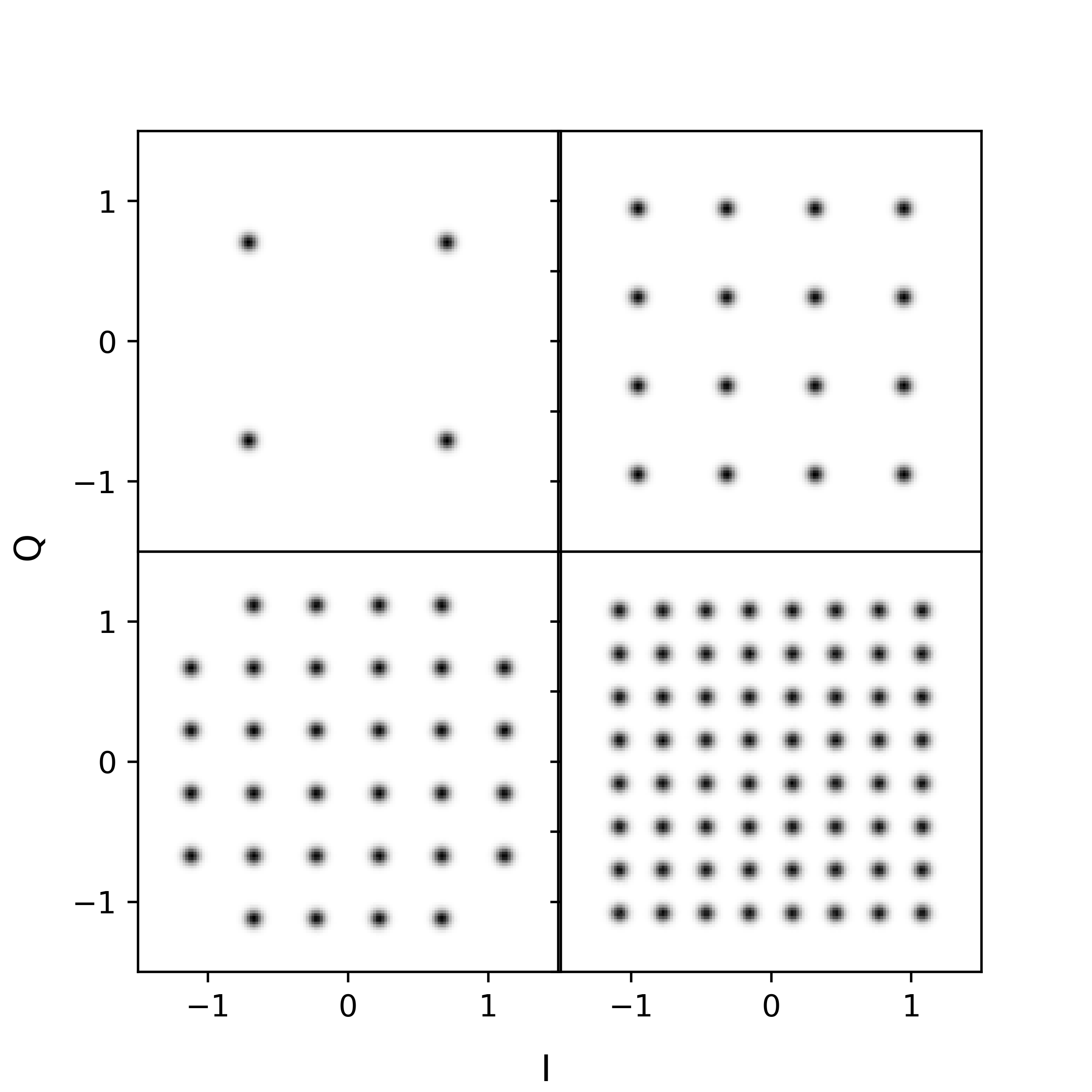}
\includegraphics[width=.85\linewidth]{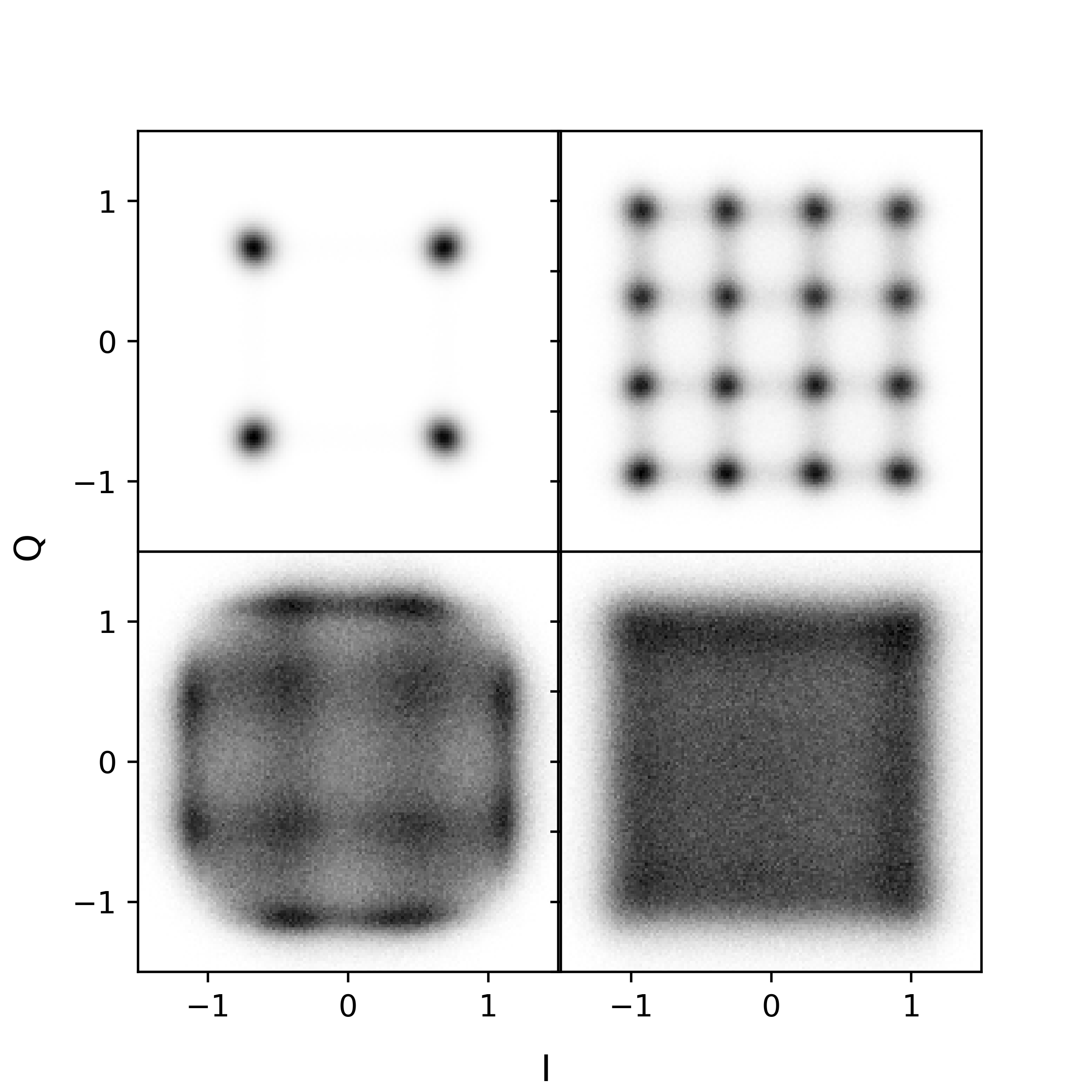}
\caption{Results for the modulation-order experiment.  Top: Target distribution constellations for 4-QAM (upper left), 16-QAM (upper right), 32-QAM (lower left), and 64-QAM (lower right).  Bottom: Generated distribution constellations.}
\label{fig:modulation_order_results}
\end{figure}

Figure~\ref{fig:modulation_order_results} shows constellation diagrams for the four modulation orders.  Performance worsened with increasing modulation order, with the constellation not clearly recovered for $M=32$ and $M=64$.  This result is expected, since increasing the modulation order both increases the total number of symbols and decreases the distance between symbols in I/Q space.  For all cases, the PSD distances ranged between $0.05$ and $0.075$, and the \ac{EVM} values for all cases were between $-$17~dB and $-$14.5~dB, both of which were consistent with the findings in the data complexity experiment. 

\subsection{Fading Channel Experiment}
\label{sec:channel_experiment}

The objective of the fading channel experiment was to evaluate the ability of \ac{STFT-GAN} to learn waveform variations due to the frequency-selective fading that arises from multipath RF propagation.  Specifically, we applied stochastic $N$-tap Rayleigh fading channel models \cite{Molisch2012} to the target distribution OFDM waveforms.  We used three channel models specified in the \ac{3GPP} cellular standard \cite[Annex B.2]{3GPP-channels}: EPA-5\,Hz, EVA-70\,Hz, and ETU-300\,Hz.  The first three letters specify a delay profile, and the frequency is the maximum Doppler frequency.

To be sensitive to frequency-dependent channel variations, the largest OFDM symbol length of $512$ was used.  In addition, the allocation size was medium, and the QAM modulation order was $M=16$.  To partially offset the nonuniform level of distortion from the different channel models, \ac{AWGN} was added such that the \ac{EVM} was $-30$\,dB, $-40$\,dB, and $-50$\,dB before application of the EPA-5\,Hz, EVA-70\,Hz, and ETU-300\,Hz channel models, respectively.

Because the 3GPP channel models are specified in terms of physical quantities, their implementation requires specification of a physical sampling rate.  We used a sampling rate of 7.68~megasamples per second, which corresponds to the 3GPP specification for a 5~MHz LTE downlink channel, where the OFDM symbol length is $512$ \cite{LTEnutshell}.  

\begin{figure}[t!]
\centering
\includegraphics[width=0.85\linewidth]{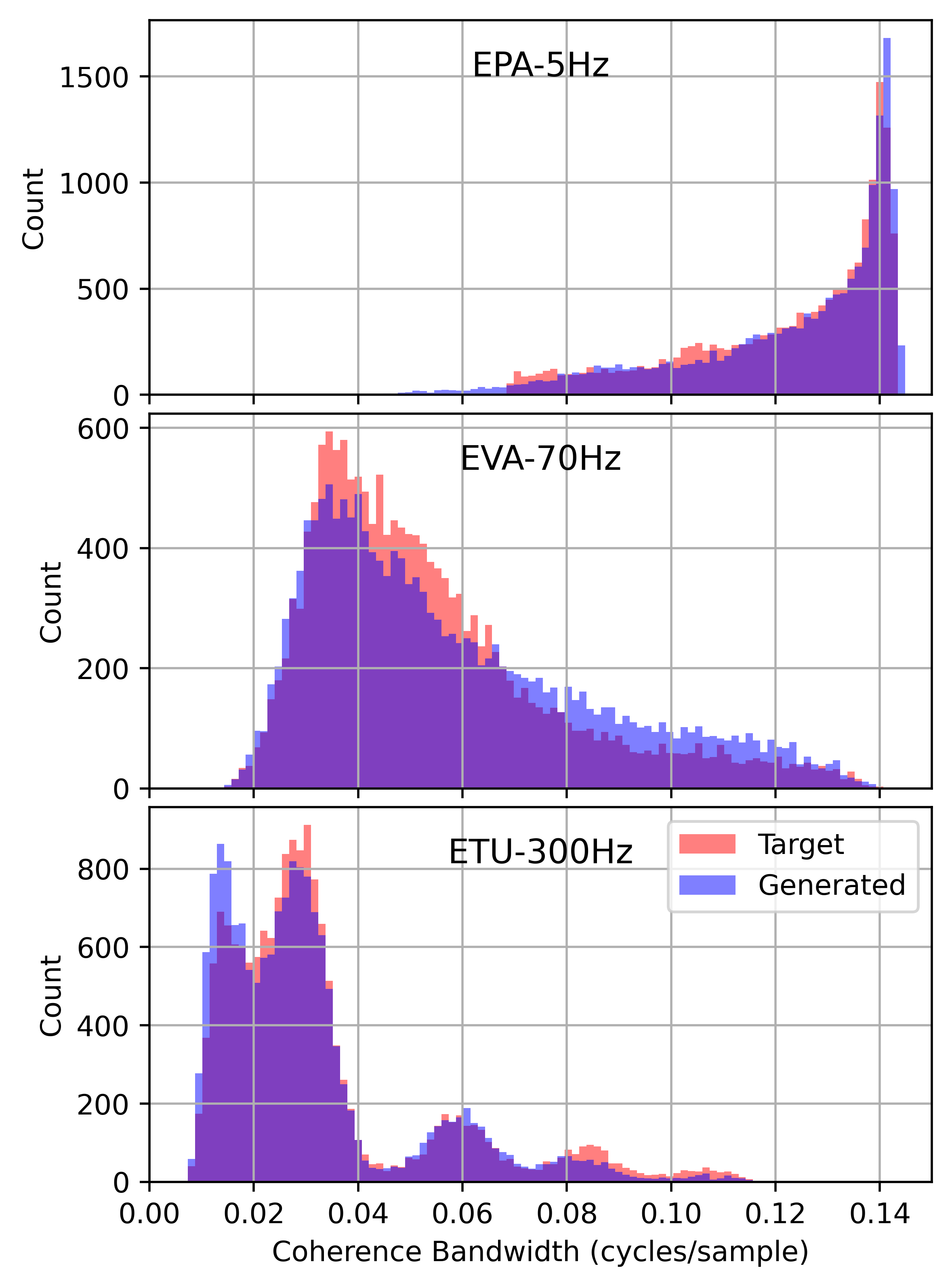}
\caption{Histograms of estimated coherence bandwidth for the fading channel experiment.}
\label{fig:Bcoh_histograms}
\end{figure}

Figure~\ref{fig:Bcoh_histograms} presents histograms of estimated coherence bandwidth for the three channels.  In each case, these plots show excellent agreement between the distributions of estimated coherence bandwidth for the generated and target data distributions.  Moreover, for all cases, the PSD distances ranged between 0.075 to 0.1, indicating strong median PSD accuracy.  

After channel equalization, the \acp{EVM} of the target and generated distributions were $-$26.8~dB and $-$12.1~dB, $-$18.1~dB and $-$11.3~dB, and $-$7.8~dB and $-$10.1~dB, for the EPA-5\,Hz, EVA-70\,Hz, and ETU-300\,Hz channel models, respectively.  These results are consistent with the findings of the data complexity experiment, indicating that there is a limit to the constellation fidelity for lower target \acp{EVM}.  Thus, while \ac{STFT-GAN} failed to achieve target constellation fidelity, it successfully learned the target PSD and channel effects quantified by the coherence bandwidth.  

\subsection{RF Impairment Experiment}
\label{sec:RF_impairment_experiment}

The aim of the RF impairment experiment was to assess the ability of \ac{STFT-GAN} to learn effects arising from imperfections in analog RF hardware.  Namely, we examined two types of RF impairments, carrier frequency offset (CFO) and I/Q imbalance; see Sec.~\ref{sec:datasets} for implementation details.  To better understand the impact of each impairment, we conducted separate subexperiments on CFO and I/Q imbalance, respectively.  For all test configurations, we used an OFDM symbol length of $256$, a QAM modulation order of $M=16$, and the large allocation size. Additionally, we added AWGN such that the $\text{EVM}=-25$~dB prior to application of the impairment transformation.  

For the CFO subexperiment, relative to a physical sampling rate of $3.84$~megasamples per second, which corresponds to the 3GPP specification for a $3$~MHz LTE downlink channel where the OFDM symbol length is $256$ \cite{LTEnutshell}, we considered two settings:  $\Delta_{\text{CFO}} = 50$\,Hz and $300$~\,Hz.  Since the subcarrier spacing in LTE is $15$~kHz, these CFO values correspond to $0.33\,\%$ and  $2\,\%$ of the subcarrier spacing, respectively.  For the I/Q imbalance subexperiment, we considered two configurations: ($\Delta_{G}$ $\Delta \phi$) = 
($10\,\%$,~$5^{\circ}$) and ($20\,\%$,~$10^{\circ}$), which we denote as ``low'' and ``high'' I/Q imbalance conditions, respectively.   

Because the CFO and I/Q impairments had minimal effects on the PSD, we do not present PSD results here.  In all cases, the PSDs for generated distributions showed strong agreement with the target distributions, similar to those observed previously.

\begin{figure}[t!]
\centering
\includegraphics[width=.85\linewidth]{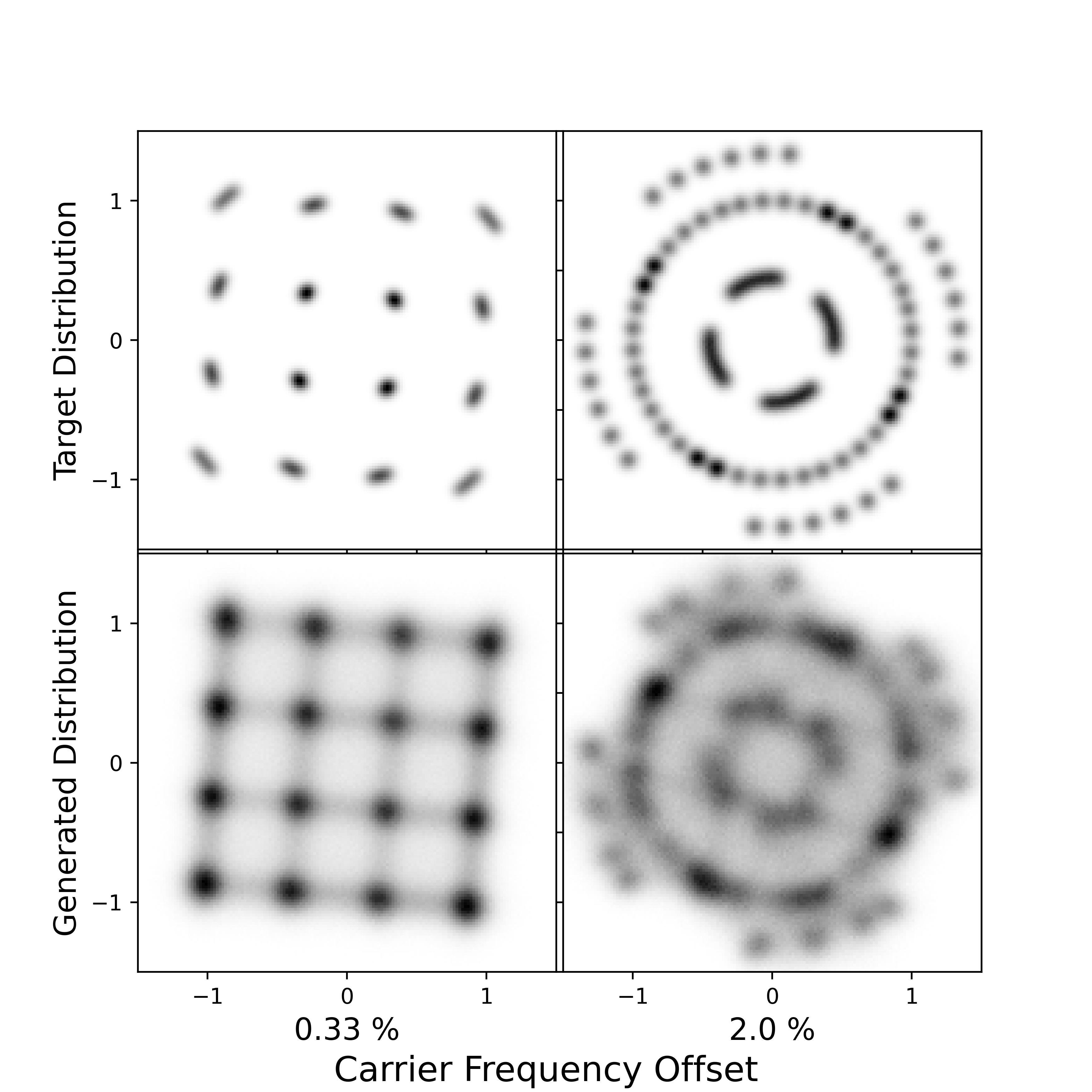}
\includegraphics[width=.85\linewidth]{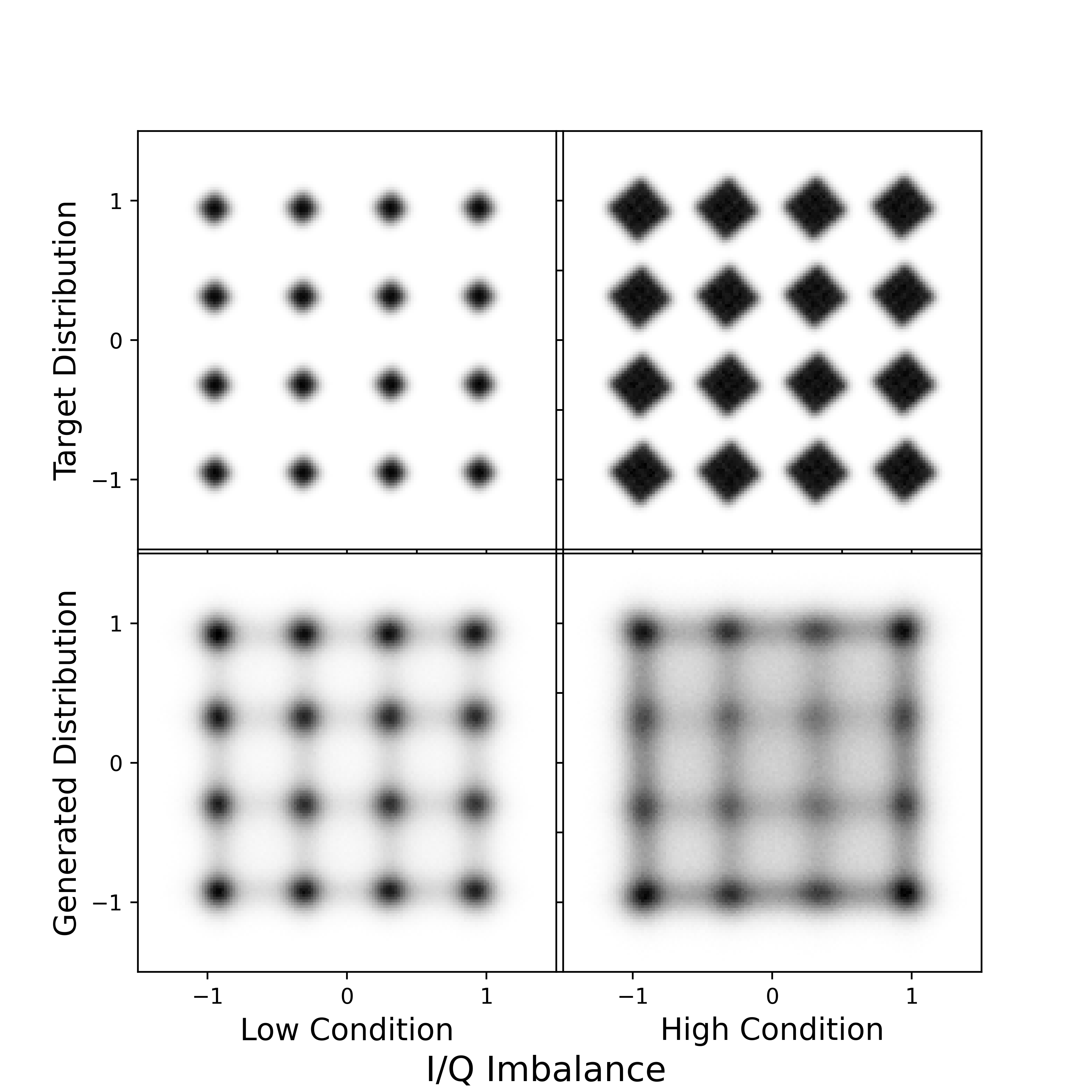}
\caption{Results for the RF impairment experiment. Top: Constellation diagrams for the carrier frequency offset (CFO) subexperiment; the CFO percentages denote frequency offset relative to subcarrier spacing.  Bottom: Constellation diagrams for the I/Q imbalance subexperiment.}
\label{fig:RF_impairment_results}
\end{figure}

Figure~\ref{fig:RF_impairment_results} presents constellation diagrams for the RF impairment experiment.  The top row shows the target distributions, and the bottom row shows the generated distributions.  The left plot, which shows the CFO experimental results, demonstrates that \ac{STFT-GAN} successfully learned several aspects of the CFO impairment, such as the symbol locations and constellation rotation.  On the other hand, the overall fidelity of the \ac{STFT-GAN} results clearly did not match the target distributions, which is consistent with the findings of the prior experiments.  The right plot, which contains the I/Q imbalance experimental results, shows that \ac{STFT-GAN} did not produce the effects caused by I/Q imbalance on the constellation.  Considering the previously observed fidelity limitations in generated constellations, this finding is not surprising, since the I/Q mismatch induces relatively subtle changes in the constellation that are below the precision of the generator. 

\section{Discussion and Conclusions}

Building on prior GAN methods, we proposed two novel GAN models, \ac{PSK-GAN} and \ac{STFT-GAN}, for I/Q OFDM time series and evaluated their performance, along with a previously published model, WaveGAN, using simulated data sets with known ground truth. Specifically, we investigated model performance with respect to increasing data set complexity over a range of OFDM parameters and conditions, including fading channels and RF impairments.  In all cases, performance evaluations were focused on metrics specific to communication signals, such as PSD and signal constellation fidelity.   

In the data complexity experiment, we found that a GAN based on an image-domain time-frequency representation, \ac{STFT-GAN}, demonstrated superior performance to two GANs that directly modeled time series, \ac{PSK-GAN} and WaveGAN.  Namely, \ac{STFT-GAN} had better PSD fidelity while maintaining lower \ac{EVM} across increasingly complex OFDM scenarios.  By contrast, both direct time-series models did not perform as well when the OFDM symbol length and allocation size increased.  Because both direct time-series models had worsening performance for longer OFDM symbol lengths and larger proportions of occupied subcarriers, they were not found to be viable candidates for further study.  Hence, the remaining two experiments focused exclusively on \ac{STFT-GAN}.

Perhaps not surprisingly, the modulation-order experiment demonstrated that increasing the QAM modulation order with the same target \ac{EVM} resulted in a decreased ability to learn the symbol constellation.  Specifically, with a target $\text{EVM}=-25$\,dB, \ac{STFT-GAN} did not learn the $M=64$ and $M=32$ constellations as well as the $M=16$ and $M=4$ constellations.  This observation is likely due to the fact that when the signal power is held constant, the density of constellation symbols in I/Q space increases with modulation order.  Future studies with higher-order modulation schemes may need to overcome this limitation.

The fading channel experiment showed that \ac{STFT-GAN} accurately learned the expected distribution of estimated coherence bandwidth for each channel type.  Moreover, \ac{STFT-GAN} learned generated distributions with median PSDs that closely matched the target distributions.  Thus, this experiment demonstrated that GANs are capable of learning signal distortions due to stochastic fading channels.     

The RF impairment experiment found that \ac{STFT-GAN} was able to successfully produce several aspects of the CFO-impaired signal constellation.  By contrast, \ac{STFT-GAN} did not produce the effects caused by I/Q imbalance on the constellation.  For both impairments, it was clear that the precision realized by \ac{STFT-GAN} was insufficient to fully capture these subtle effects on the signal constellation.  

In all experiments, \ac{STFT-GAN} achieved excellent PSD accuracy, including in regions outside the set of active subcarriers.  Moreover, it reliably learned the time-domain cyclic prefix location.  On the other hand, while \ac{STFT-GAN} learned the QAM signal constellation in many cases, it did not achieve the desired constellation fidelity as measured by \ac{EVM}.  This finding indicates a limitation of the \ac{STFT-GAN} model and training protocol employed in this work.  The potential for additional model optimization, \emph{e.g.}, deeper models and longer training, to improve constellation fidelity could be explored in future studies.

Overall, these findings indicate those use cases that are feasible for \ac{STFT-GAN} as well as those where additional advances are needed.  Namely, \ac{STFT-GAN} is likely not a good candidate for end-to-end communication system modeling, unsupervised encoding/decoding, or bit-error-rate testing where it is important to achieve accurate constellation fidelity.  However, \ac{STFT-GAN} holds promise for waveform generation in experimental interference testing, \emph{e.g.}, Ref. \cite{Young2021}, where it is crucial to have good PSD fidelity to reflect in-band and out-of-band power characteristics and to capture time-frequency channel effects, but where an accurate signal constellation is not necessary.  In future work, we plan to build on this investigation by developing generative models using real-world recordings of RF emissions for application to laboratory-based interference studies with real systems.   

\section{Appendix: WaveGAN}

A state-of-the-art \ac{GAN} that directly models time series with \acp{CNN} is WaveGAN \cite{Donahue2019}.  In Sec.~\ref{sec:main_experiment}, we compared an implementation of WaveGAN, described below, against other \ac{GAN} models for \ac{OFDM} data.  The original WaveGAN model produces single-channel audio waveforms of length $16\,384$, using an architecture based on a 1-D flattened version of the popular DCGAN model for images \cite{Radford2015}.  WaveGAN attempts to widen its convolutional receptive field by modifying the DCGAN 5x5 kernels with 2x2 strides to 1-D kernels of length 25 with strides of 4 \cite{Donahue2019}.  Also, in contrast to DCGAN, WaveGAN has no normalization layers, adds a fully connected layer to both models, and is trained with \ac{WGAN}-GP loss. 

In addition, WaveGAN adds a novel layer to the discriminator, called phase shuffle, which performs a random circular shift of a convolutional layer's output activation.  Donahue \emph{et al.} \cite{Donahue2019} found optimal results when the random circular shift was between $-2$ and $2$ time steps.  The aim of the phase shuffle operation is to make the model insensitive to the input waveform's phase, preventing so-called ``checkerboard'' artifacts, which manifest as spikes in the power spectrum \cite{odena2016deconvolution}.

Our slightly modified implementation of WaveGAN is shown in Tables~\ref{tab:WaveGAN-gen} and \ref{tab:WaveGAN-disc}, where $f=1$, $2$, or $4$ for OFDM symbol lengths of 128, 256, and 512, respectively, and $n$ is the batch size.  Like the original WaveGAN model, our implementation used five convolutional layers for the generator and discriminator, but the dense layer was modified to support the lengths of our synthetic OFDM waveforms.  

We attempted to implement phase shuffle for layers for which it was compatible, \emph{i.e.}, layers with output dimension more than 4.  However, we observed training instabilities with phase shuffle and therefore did not include it in our implementation.  Note that the lack of phase shuffle may explain the spikes in the power spectra observed in Fig.~\ref{fig:PSD_examples_main_exp}.

Following the original WaveGAN training implementation \cite{Donahue2019}, we trained the model with a 5:1 update ratio between the generator and discriminator and the Adam optimizer with $\beta_1=0.5$ and $\beta_2=0.9$.  Otherwise, we followed the same training protocol described in Sec. ~\ref{sec:training}.  Namely, the learning rate was $\alpha = 10^{-4}$ for the generator and discriminator, the batch size was $128$, and the model was trained for $500$ epochs.  Also, all target distributions were scaled between [$-1$, $1$] using feature-based min-max scaling.  

\begin{table}[t!]
  \centering
    \caption{WaveGAN generator architecture [$f$ = 1, 2, 4].}
    \label{tab:WaveGAN-gen}
    \begin{tabular}{l|c|r} 
      \textbf{Operation} & \textbf{Filter Shape} & \textbf{Output Shape}\\
      \hline
      $z$ $\sim$ Uniform(-1, 1) &  & ($n$, 100)\\
      Dense & (100, 1024$f$) & ($n$, 1024$f$)\\
      Reshape & & ($n$, 1024, $f$)\\
      ReLU & & ($n$, 1024, $f$)\\
      Transpose Conv1-D (stride=4) & (25, 1024, 512) & ($n$, 512, 4$f$)\\
      ReLU & & ($n$, 512, 4$f$)\\
      Transpose Conv1-D (stride=4) & (25, 512, 256) & ($n$, 256, 16$f$)\\
      ReLU & & ($n$, 256, 16$f$)\\
      Transpose Conv1-D (stride=4) & (25, 256, 128) & ($n$, 128, 64$f$)\\
      ReLU & & ($n$, 128, 64$f$)\\
      Transpose Conv1-D (stride=4) & (25, 128, 64) & ($n$, 64, 256$f$)\\
      ReLU & & ($n$, 64, 256$f$)\\
      Transpose Conv1-D (stride=4) & (25, 64, 2) & ($n$, 2, 1024$f$)\\
      Tanh & & ($n$, 2, 1024$f$)\\
    \end{tabular}
\end{table}

\begin{table}[t!]
  \centering
    \caption{WaveGAN discriminator architecture [$f$ = 1, 2, 4].}
    \label{tab:WaveGAN-disc}
    \begin{tabular}{l|c|r} 
      \textbf{Operation} & \textbf{Filter Shape} & \textbf{Output Shape}\\
      \hline
      $x$ $\sim$ $G$($z$) & & ($n$, 2, 1024$f$)\\
      Conv1-D (stride=4) & (25, 2, 64) & ($n$, 64, 256$f$)\\
      LReLU($\alpha=0.2$) & & ($n$, 64, 256$f$)\\
      Conv1-D (stride=4) & (25, 64, 128) & ($n$, 128, 64$f$)\\
      LReLU($\alpha=0.2$) & & ($n$, 128, 64$f$)\\
      Conv1-D (stride=4) & (25, 128, 256) & ($n$, 256, 16$f$)\\
      LReLU($\alpha=0.2$) & & ($n$, 256, 16$f$)\\      
      Conv1-D (stride=4) & (25, 256, 512) & ($n$, 512, 4$f$)\\
      LReLU($\alpha=0.2$) & & ($n$, 512, 4$f$)\\
      Conv1-D (stride=4) & (25, 512, 1024) & ($n$, 1024, $f$)\\
      LReLU($\alpha=0.2$) & & ($n$, 1024, $f$)\\
      Reshape & & ($n$, 1024$f$) \\
      Dense & (1024$f$, 1) & ($n$, 1) \\
    \end{tabular}
\end{table}

\begin{acronym}
\acro{3GPP}{Third-Generation Partnership Project}
\acro{AWGN}{additive white Gaussian noise}
\acro{BER}{bit error rate}
\acro{CNN}{convolutional neural network}
\acro{COLA}{constant-overlap-add}
\acro{COTS}{commercial off-the-shelf}
\acro{dB}{decibels}
\acro{DMRS}{demodulation reference signal}
\acro{DFT}{discrete Fourier transform}
\acro{EVM}{error vector magnitude}
\acro{FFT}{fast Fourier transform}
\acro{GAN}{generative adversarial network}
\acro{ICI}{intercarrier interference}
\acro{I/Q}{in-phase and quadrature}
\acro{ISI}{intersymbol interference}
\acro{LSTM}{long short-term memory}
\acro{LTE}{long-term evolution}
\acro{OFDM}{orthogonal frequency-division multiplexing}
\acro{PSD}{power spectral density}
\acro{QAM}{quadrature amplitude modulation}
\acro{PSK-GAN}{progressively scaled kernel GAN}
\acro{RF}{radio frequency}
\acro{RMS}{root mean square}
\acro{RNN}{recurrent neural network}
\acro{SNR}{signal-to-noise ratio}
\acro{STFT}{short-time Fourier transform}
\acro{STFT-GAN}{short-time Fourier transform GAN}
\acro{TCN}{temporal convolutional network}
\acro{UE}{user equipment}
\acro{VAE}{variational autoencoder}
\acro{WLAN}{wireless local area network}
\acro{WGAN}{Wasserstein GAN}
\acro{WSSUS}{wide-sense stationary uncorrelated scatterers}
\end{acronym}
\bibliographystyle{IEEEtran}
\bibliography{references.bib}
\end{document}